\documentclass[12pt]{iopart}
\expandafter\let\csname equation*\endcsname=\relax 
\expandafter\let\csname endequation*\endcsname=\relax 

\usepackage{graphicx}
\usepackage{amssymb}
\usepackage{amsmath}
\usepackage[
	  colorlinks=true,
      linkcolor=blue,  
      urlcolor=black,    
      filecolor=blue,     
      citecolor=red,
	  linktocpage=true,
      pdfstartview=FitV,
      bookmarksopen=true,
      pdftitle={BootAmps}]{hyperref}

\usepackage{etoolbox}

\usepackage{tikz}
\usetikzlibrary{arrows,shapes}
\usetikzlibrary{trees}
\usetikzlibrary{patterns}
\usetikzlibrary{matrix,arrows} 				
\usetikzlibrary{positioning}				
\usetikzlibrary{calc,through}				
\usetikzlibrary{decorations.pathreplacing}  
\usepackage{pgffor}							

\usetikzlibrary{decorations.pathmorphing}	
\usetikzlibrary{decorations.markings}
\tikzset{
    vector/.style={decorate, decoration={snake}, draw},
	provector/.style={decorate, decoration={snake,amplitude=2.5pt}, draw},
	antivector/.style={decorate, decoration={snake,amplitude=-2.5pt}, draw},
        smallvector/.style={decorate, decoration={snake,amplitude=1.5pt,post length=0.5mm}, draw},
    fermion/.style={draw=black, postaction={decorate},
        decoration={markings,mark=at position .55 with {\arrow[draw=black]{>}}}},
    fermionbar/.style={draw=black, postaction={decorate},
        decoration={markings,mark=at position .55 with {\arrow[draw=black]{<}}}},
    fermionnoarrow/.style={draw=black},
    gluon/.style={decorate, draw=black,
        decoration={coil,amplitude=4pt, segment length=5pt}},
    scalar/.style={dashed,draw=black, postaction={decorate},
        decoration={markings,mark=at position .55 with {\arrow[draw=black]{>}}}},
    scalarbar/.style={dashed,draw=black, postaction={decorate},
        decoration={markings,mark=at position .55 with {\arrow[draw=black]{<}}}},
    scalarnoarrow/.style={dashed,draw=black},
    electron/.style={draw=black, postaction={decorate},
        decoration={markings,mark=at position .55 with {\arrow[draw=black]{>}}}},
    bigvector/.style={decorate, decoration={snake,amplitude=4pt}, draw},
    arrow/.style={draw=black, postaction={decorate},
        decoration={markings,mark=at position 1 with {\arrow[draw=black]{>}}}},
}

\tikzstyle{block} = [draw, rectangle, 
    minimum height=3em, minimum width=6earticlem]

\def\be{\begin{equation}}
\def\ee{\end{equation}}
\def\bea{\begin{eqnarray}}
\def\eea{\end{eqnarray}}
\def\<{\langle}
\def\>{\rangle}
\def\pa{\partial}   
\def\a{\alpha}   
   
\newcommand{\eps}{\epsilon}
\newcommand{\lra}{\leftrightarrow}
\newcommand{\reef}[1]{(\ref{#1})}

\makeatletter
\newcommand{\mainmatter}{%
  \setcounter{footnote}{0}%
  \patchcmd{\@makefntext}{\fnsymbol}{\arabic}{}{}%
  \patchcmd{\@thefnmark}{\fnsymbol}{\arabic}{}{}%
  \def\@makefnmark{\textsuperscript{\arabic{footnote}}}%
}
\makeatother

\begin{document}
\title[Bootstrap and Amplitudes]{Bootstrap and Amplitudes: A Hike in the Landscape of Quantum Field Theory}

\author{Henriette Elvang}

\address{Randall Laboratory of Physics, Department of Physics,\\
and Leinweber Center for Theoretical Physics (LCTP),\\
	   University of Michigan, Ann Arbor, MI 48109, USA}
\address{LCPT-20-17}	   
\ead{{\href{mailto:elvang@umich.edu}{elvang@umich.edu}}}
\date{}

\begin{abstract}
This article is an introduction to two currently very active research programs, the Conformal Bootstrap  and Scattering Amplitudes. Rather than attempting full surveys, the emphasis is on common ideas and methods shared by these two seemingly very different programs. 
In both fields, mathematical and physical constraints are placed directly on the physical observables  in order to explore the landscape of possible consistent quantum field theories.  We give explicit examples from both programs: the reader can expect to encounter boiling water, ferromagnets, pion scattering, and emergent symmetries on this journey into the landscape of local relativistic quantum field theories. The first part is written for a general physics audience. The second part includes further details, including a new on-shell bottom-up reconstruction of the $\mathbb{CP}^1$ model with the Fubini-Study metric arising from re-summation of the $n$-point interaction terms derived from amplitudes.

The presentation is an extended version of  a colloquium given at the Aspen Center for Physics in August 2019.

\end{abstract}

\maketitle

\mainmatter

\tableofcontents

\title[Bootstrap and Amplitudes]{}

\newpage

\section{Introduction}
\label{sec:Intro}

What do boiling water and ferromagnetics have in common? At first sight: not much. However, near the critical point in their phase diagrams, water and ferromagnets exhibit a similar behavior: various physical quantities scale in the same way, with exactly the same critical exponents. This is an example of universality. Despite their widely different microscopics,  systems in the same universality class can be described at the critical point by the same scale-invariant theory.

Another very different class of physics is explored by scattering experiments, such as the Large Hadron Collider (LHC) at CERN. In order to compare with experimental data, particle theorists compute scattering amplitudes that encode the probability that a given initial state interacts and scatters to a particular final state.   A simple example is the scattering of pions $\pi$, a light hadron associated with the strong nuclear force. For a process $\pi \pi \to \pi\pi$, the amplitude $A_4(\pi \pi \to \pi\pi)$ encodes the probability of the process as a function of the center of mass energy and the scattering angle. Specifically, the (differential) scattering cross-section is proportional to a phase-space integral over the norm-squared of the amplitude $|A_4|^2$. The theory that describes interacting massless pions is absolutely not scale-invariant and it is therefore completely different from the type of theory that describes the critical point  of boiling water and ferromagnets. 

Despite their obvious differences,  boiling water, ferromagnets, and pion scattering   are part of a   vastly broader class of physical systems that are explored using a set of powerful methods in modern theoretical physics. The basic idea is to ``bootstrap" the physical observables directly from physical and mathematical consistency constraints rather than calculating them from detailed microscopic descriptions. 
One then uses the observables --- subject to desired properties and symmetries --- to learn about the landscape of possible theoretical models that can give rise to such observables.  
A specific goal is to understand the structure of {\em quantum field theories (QFTs)}. 

QFT is a mathematical framework for theoretical physics. It has a plethora of applications and direct experimental relevance. QFT is relevant for particle physics, condensed matter systems, string theory, gravity, gravitational waves, and beyond. There is not {\em one} quantum field theory but {\em many}. Some QFTs describe particles that are weakly interacting and one can use Lagrangian techniques to study them. Other QFTs are always strongly coupled; in those cases words such as ``particles" and their ``interactions" are not useful and they may have no  Lagrangian descriptions. Some QFTs describe physics that depends heavily on the energy scale (or length scale) while other QFTs do not care a whit about scale. 

The subject of QFT is incredibly rich. QFTs describe the critical points of water and ferromagnets as well as the scattering of pions. The set of consistent QFTs can be thought of as a landscape: an abstract landscape that is so vast and complex and interesting that theorists constantly venture into its unknowns to explore and discover new features, new connections, and new properties. 

It can be hazardous to venture out on a hike into unknown terrain, so we consult maps in order to know the local topographical features of the landscape; such as the beautiful Rocky Mountains.\footnote{This article is based  on a colloquium given August 8, 2019, at the Aspen Center for Physics, Aspen, Colorado, USA.}  
The peaks of the mountains, the valleys, and the saddlepoints are the most prominent features and they guide our choice of path. 
Likewise, we wish to map out the landscape of QFTs. There are special places in the QFT landscape that can help us understand it better and navigate it. It is very useful to determine these special QFTs and understand their properties. Examples of such  special points in the QFT landscape are known as {\em conformal field theories (CFTs)}: the CFTs are the metaphorical peaks, ridges, and valleys of the QFT landscape. 

Two modern approaches to explore the landscape of QFTs are:
\begin{itemize}
\item  the Conformal Bootstrap Program focused on the CFTs, and
\item  the Scattering Amplitudes Program.
\end{itemize}
 The goal of this article is to give colloquium-level introductions to these two highly active research areas and describe how they share a common approach to physics that leads to powerful and novel results. It is my hope that this will be useful for researchers in other fields of physics and math as well as for students. For those with more background in QFT, I have included two sections with technical details beyond the colloquium-level because I wanted to illustrate the ideas concretely and explicitly. 

\vspace{2mm}
\noindent {\bf Overview.}
The presentation has two main parts: the first part --- Sections \ref{s:watermagnet} through \ref{s:introbootstrap} --- is intended for a general physics audience with no prior knowledge of the subjects. The second part  --- Sections \ref{s:ampbootex} and \ref{s:confboot}  --- provides  technical details that put more equations behind the words in the first part.

We begin with the description of the critical points of water and ferromagnets in Section \ref{s:watermagnet}: we discuss critical exponents and scale invariance. It has been proposed that a 3d conformal field theory describes the physics at the critical point. Before exploring that 3d theory further, we illustrate in Section \ref{s:relQFT} the  richness of the landscape of 4d relativistic quantum field theories by describing a few examples of QFTs and we then introduce conformal field theories. In Section  \ref{s:amps}, we introduce the modern amplitudes program with focus on the ideas of using scattering amplitudes to explore the landscape of QFTs. Section \ref{s:introbootstrap} offers an introduction to the conformal bootstrap program with particular emphasis on what it teaches us about the critical points of boiling water and ferromagnets. 

The presentations in Sections \ref{s:amps} and \ref{s:introbootstrap} attempt to avoid the full technical detail, but for those who want to see more, please see Sections  \ref{s:ampbootex} and \ref{s:confboot}. In particular, Section \ref{s:ampbootex} provides the full details of how to bootstrap a scalar model from very simple assumptions about the behavior of the scattering processes and we shall see how a global symmetry emerges from the construction. We show  how the Lagrangian interaction terms can be reconstructed from the bootstrapped amplitudes and that they can be re-summed to the Fubini-Study metric. Section \ref{s:confboot} presents technical detail about the conformal bootstrap setup and as an example it is shown how the crossing relations requires an interacting 4d CFT to have an infinite number primary operators. We conclude in Section \ref{s:conclude} with very brief closing remarks.

\section{Water \& Magnets}
\label{s:watermagnet}

\begin{figure}[t]
\begin{center}
    \begin{tikzpicture}[scale=0.4, line width=1 pt]
    	\draw (0,0)--(19,0);
	\draw (0,0)--(0,16);
	\draw[orange] (0,1.5) .. controls (3,2) and (13, 3) .. (18, 13);
	\draw[orange] (3.3,2.1) .. controls (2.3,6) .. (1.5, 15.5);
	\draw[dotted,gray] (0,13)--(18, 13);
	\draw[dotted,gray] (18,0)--(18, 13);
	\draw[dotted,gray] (0,2.1)--(3.3,2.1); 
	\draw[dotted,gray] (3.3,0)--(3.3,2.1);
	\draw[dotted,gray] (0,5)--(10.8,5);
	\draw[dotted,gray] (10.8,0)--(10.8,5);
	\draw[dotted,gray] (2.6,0)--(2.6,5); 
	\fill[orange] (3.3,2.1) circle (1.5ex); 
	\fill[orange] (18, 13) circle (1.5ex); 
	\fill[orange] (10.8,5) circle (1.5ex); 
	\fill[orange] (2.6,5) circle (1.5ex); 
	\node[rotate=90] at (-3,7.5) {\small pressure in atm};
	\node at (16,-2) {\small temperature in $^\circ$C};
	\node at (18, 13.7) {\tiny critical point};
	\node at (5.4,1.8) {\tiny triple point};
	\node at (14.1,4.7) {\tiny normal boiling point};
	\node at (5.9,5.5) {\tiny normal freezing point};
	\node at (2.6,-0.6) {\footnotesize $0$};
	\node at (4,-0.6) {\footnotesize $0.01$};
	\node at (10.8,-0.6) {\footnotesize $100$};
	\node at (18,-0.6) {\footnotesize $374$};
	\node at (-1.7,2.1) {\footnotesize $0.0060$};
	\node at (-0.7,5) {\footnotesize $1$};
	\node at (-1.7,13) {\footnotesize $217.75$};
	\node[rotate=90,gray] at (1, 9) {\footnotesize solid};
	\node[gray] at (9, 10) {\footnotesize liquid};
	\node[gray] at (14, 3) {\footnotesize gas};
    \end{tikzpicture}
\end{center}
\caption{\label{waterphases} Sketch of the phase diagram of water. Across the orange lines, the changes of phase require latent heat and are first order phase transitions. At the critical point, the phase transition becomes second order; at this point the system  becomes scale invariant. It is the scale invariant model at the critical point that is of interest here.}
\end{figure}
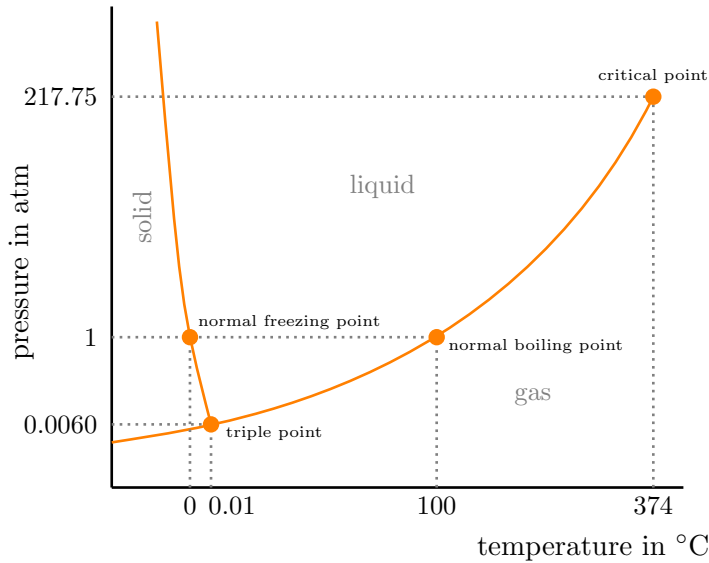

Consider the phase diagram of water in Figure \ref{waterphases}. Under normal conditions of pressure at about 1\,atm, water freezes at $0^\circ$C and boils at $100^\circ$C, so as the temperature is varied at constant pressure, water exhibits three phases: solid, liquid, and gas. As is well-known by people in mountainous regions and students in thermodynamics classes, the boiling point is lower at higher altitude. In Aspen, at about 8,000\,ft = 2440\,m, the air pressure drops to around 0.75\,atm and the boiling point of water is $92^\circ$C. So it takes a little longer to boil your pasta al dente. 

The familiar solid-liquid and liquid-gas phase transitions of water involve latent heat and are called {\em first order phase transitions}. As  pressure  increases, the boiling point of water goes up and at high enough pressure, $p > 217$\,atm, the phases of liquid and gas are no longer distinguishable. The liquid-gas transition curve in the phase diagram ends at a point called the {\em critical point} with $p_c \sim 217$\,atm and $T_c \sim 374^\circ$C. As the critical point is approached, the latent heat needed to transition between liquid and gas goes to zero and at the critical point the phase transition becomes {\em continuous} (also called {\em second order}). 

OK, so what? Well, near the critical point, something special happens to the correlation length $\xi$ in the system. The correlation length says something about how strongly coupled disparate parts of the system are to each other. As $T \to T_c$, the correlation length diverges as 
\be 
  \label{xi}
   \xi \sim |(T-T_c)/T_c|^{-\nu} \,.
\ee
Points at large spatial separation $r$ are correlated with strength $e^{-r/\xi}$, so an infinite correlation length, $\xi \to \infty$, means that every part of the system couples with equal strength to every other part. Not just nearest-neighbor friendliness here, everybody is coupled to everybody else. Moreover, when $\xi\to\infty$  there is no distinguishing scale in the system: it has become {\em scale invariant}. Physically, the phenomenon of scale invariance can be seen as critical opalescence: at the critical point the liquid becomes milky in appearance as the correlation lengths that govern the fluctuations in the system become of the same order as the wavelength of visible light so it scatters and the substance looks cloudy.  

The number $\nu$ in equation \reef{xi} is an example of a critical exponent. These exponents characterize the approach to the critical point and play an important role for critical systems. For  critical point of a liquid-vapor transition, the value of $\nu$  is
\be
\nu \approx 0.63\,. 
\ee

Now let us switch gears and discuss a physically very different system, namely ferromagnets. They exhibit critical behavior at the Curie temperature $T_c$, which separates the ordered ferromagnetic phase at $T<T_c$ from the disordered non-magnetic phase at  $T>T_c$. As an example, the Curie temperature of iron is $T_c =1043$\,K = $770^\circ$C; for reference, iron melts at $1811$\,K =$1538^\circ$C. At the critical point set by the Curie temperature, the correlation length between dipoles in the ferromagnet diverges just as in equation \reef{xi}. And here comes the best part: {\em the critical exponent $\nu$ for ferromagnets takes the same value as for water $\nu \approx 0.63$.} This is quite amazing: these are different physical phenomena whose microscopics are totally unrelated. Nonetheless, at their critical points, these two systems --- water and ferromagnets --- behave alike. This is an example of {\em universality}. 

There is a theoretical model, the  3d Ising model, that describes both water and ferromagnets near their critical points. For the ferromagnet, this is a model in which each site of a 3d lattice 
can either be spin up or downs. For water, replace spin up/down with occupation number: site has a molecule or not. Block spin techniques allow one to study such a lattice model at greater and greater length scales, meaning lower and lower energies. In the deep infrared, one approaches a fixed point of scale invariance (called the critical 3d Ising model) and finds that the correlation length diverges with a critical exponent $\nu \approx 0.63$. 

It appears that the value of $\nu \approx 0.63$ has not been measured directly experimentally for the critical point of regular water H$_2$O, but it can be inferred from other measurements of critical exponents. There are, however, multiple other direct measurements of $\nu$ in other systems: small angle neutron scattering in heavy water D$_2$O \cite{Sullivan_2000},  light-scattering experiments in an electrolyte solution \cite{Sengers2009} as well as other systems in the same 3d Ising universality class, see for example Table 7 in \cite{Pelissetto:2000ek}. 

Now you may have started reading this article in the hope of learning about the landscape of QFTs and instead you have gotten an earful about the phase diagrams of water and ferromagnets. Fear not, there is a purpose behind   the madness. Polyakov conjectured that \cite{Polyakov:1970xd} that at critical points, the symmetries of the system are enhanced to conformal symmetry: they can be described by conformal field theories. 
We have started out with the critical continuous phase transitions in water and ferromagnets because these are real-world examples of the power of quantum field theory and a case where the conformal bootstrap techniques are directly applicable. The point here is that by studying the 3d Ising model (henceforth referring to it with implicit understanding that it is at the fixed point) as a conformal field theory using the techniques we describe in Section \ref{s:introbootstrap}, one can extract, to an incredible precision, information about the critical exponents of the system, such as $\nu$.

\section{Field Theories: A Brief Survey}
\label{s:relQFT}

In this article we study  {\em relativistic} QFT.\footnote{Non-relativistic QFT is an important subject too that is highly relevant in condensed matter contexts. The examples of water and ferromagnetics are non-relativistic, but at the critical points the symmetries are expected to be enhanced, as discussed at the end of Section \ref{s:watermagnet}.} This means that we consider QFTs that are invariant under Poincar\'e symmetry: spatial rotations, Lorentz boosts, and spacetime translations. Moreover, we assume that the theories we study are local and unitary. Loosely speaking locality means that there is no action at a distance. Technically, this means that local fluctuations can be described in terms of local operators that depend on a single point of spacetime. Locality manifests itself on the physical observables in terms of what kind of singularity structures they are allowed to have. We discuss this further in Section \ref{s:amps}.

As stated in the Introduction, there is not just one QFT but a vast and rich landscape of QFTs. To illustrate how different QFTs can be --- and to set the stage for the later discussion --- I will now briefly discuss some examples of relativistic QFTs.

\subsection{Examples of Relativistic QFTs}
\label{s:exQFTs}

Here follows some key examples of 4d quantum field theories:
\begin{itemize}
\item {\em Quantum Electrodynamics (QED)} describes the interaction of photons and electrons/positrons. The classical equations of motion are Maxwell's equations for the photons and the Dirac equation for the electrons/positrons. At the quantum level, the strength of the coupling of photons to electrons/positrons depends on the energy scale. At atomic-level energies the effective coupling, the fine structure constant $\alpha$, is $\alpha \approx 1/137$. However, at energies around the mass of  the weak force mediators $W^\pm$ and $Z$, which is about $90$ times the proton mass, the strength of the coupling increases to around $\alpha \approx 1/127$. Thus the coupling runs with scale: it approaches zero at very low energies where the theory becomes trivial, but at high energies it becomes stronger and perturbation theory breaks down. 
\item {\em Quantum Chromodynamics (QCD)} describes the strong nuclear force. More precisely, QCD is the theory that describes gluons, $N_f$ flavors of quarks (with $N_f=6$ in Nature for the $d$, $u$, $s$, $c$, $b$, and $t$ quarks), and their interactions. The dynamics of gluons is captured by {\em Yang-Mills theory}. The coupling strength $\alpha_s$ of QCD also depends on the energy scale. Famously, QCD in our world behaves oppositely to QED in that it becomes free (coupling goes to zero) at high energies while it is strongly coupled and confining at low energies. 

\item {\em The Standard Model of Particle Physics} combines QED with three generations of leptons (electrons, muons, taus, and their antiparticles, as well as the neutrinos), QCD with six flavors of quarks, the electroweak force, and the Higgs mechanism  to give an incredibly successful quantum field theory in which one can calculate physical observables to high precision and compare with experimental data. An example is the multi-digit precision agreement in the measurements of the electron and muon Land\'e $g$-factor. The Higgs mechanism and spontaneously broken symmetry are key ingredients in the Standard Model and the discovery of the Higgs boson announced in 2012 was a tremendous success of the both long-term experimental perseverance and the power of theoretical studies of quantum field theories.  

\vspace{0.8mm}
In the Standard Model, the couplings also run with energy scale. The electromagnetic force and the weak nuclear force unify at about 246 times the proton mass. Grand Unification models seek to unify the electroweak force with the strong force at an even higher scale. 

\item {\em Gravity}  is not included in the Standard Model. However, it is successfully described by its own field theory, namely General Relativity. As a field theory, General Relativity differs from those discussed above in that the gravitational coupling $\kappa = G^{1/2}$ is dimensionful. Specifically, Newton's constant $G$ has dimension of (mass)$^{-2}$. This is in contrast with the dimensionless fine structure constant $\alpha$ of QED. As a consequence, the effective dimensionless coupling in gravity is $E \kappa = E \sqrt{G}$, where $E$ is the energy scale in the particular problem. Thus gravity is a weak force (i.e.~perturbative) only at energy scales much smaller than $\kappa^{-1} = G^{-1/2} \sim M_\text{Planck} \sim 10^{19}$ GeV. In a sense this is like QED in that gravity is weak at low-energies and strong at high energies. But gravity is actually much more complicated than QED. QED is a renormalizable theory meaning that it is a sensible predictive theory at the quantum level. Gravity, on the contrary, is non-renormalizable and therefore, broadly speaking, it is non-predictive at the quantum level at energies approaching the Planck scale.  

\vspace{0.8mm}
It is useful to think of General Relativity is as a  {\em low-energy effective field theory}: it gives a description of gravity that makes sense only at sufficiently low energies $E \ll M_\text{Planck}$. Within this regime of validity it is, as we know well, a highly successful theory of gravitational phenomena. At energies above the Planck scale, it needs a UV completion in order to make sense.  String theory is a theoretical framework that, among many other properties, is the most promising candidate for a theory of quantum gravity that gives a UV completion of General Relativity. 

\item {\em Effective Field Theories (EFTs)} describe physical phenomena in an expansion in some small parameter(s). In the context here, we focus on 
low-energy EFTs. The idea is to work in a low-energy regime, where powers of energy-momentum are suppressed by a particular ``cut-off'' scale $\Lambda_\text{UV}$ above which the expansion in small $E/\Lambda_\text{UV}$ is no longer valid. In terms of a Lagrangian, derivatives are (via Fourier transform) directly counting powers of energy-momentum, so in an EFT one includes higher-derivative terms with increasing suppression by powers of  $1/\Lambda_\text{UV}$. This means that the $1/\Lambda_\text{UV}$-expansion becomes a derivative expansion. The principle of EFTs is to include all Lagrangian terms that are allowed by symmetries of the system up to a given order in $1/\Lambda_\text{UV}$. The arbitrary coefficients in this expansion parameterize  the (potentially unknown) UV physics. 

\vspace{0.8mm}
As an example, General Relativity is the leading 2-derivative interactions of a gravitational EFT in which higher-derivative corrections are suppressed by inverse powers of the Planck mass.  The historical successes of General Relativity --- as well as the frequent use you make of it via the GPS built into your smart phone --- tells you that effective field theories are incredibly useful. 

\vspace{0.8mm}
Physics beyond the standard model, such as proton decay or dark matter, is often encoded in terms of EFTs. The scale of the EFT is then associated with the scale of the new physics at higher energies.

\item {\em Non-Linear Sigma Models (NLSM)} describe scalar fields that take values in some manifold, called a target space, and the model inherits the symmetries from this target space. An important class NLSM are the EFTs that govern the low-energy dynamics of massless Goldstone bosons arising from spontaneously broken global symmetries. In these cases, the scale built into the EFT is associated with the symmetry-breaking scale.

\vspace{0.8mm}
As an example, the Standard Model has an approximate chiral symmetry that is spontaneously broken at a scale of about $1$\,GeV (i.e.~about the mass of the proton). The symmetry breaking gives rise to Goldstone bosons that are 
 identified as the pions.  While Goldstone bosons are exactly massless, pions in nature are not; they have low mass (135-140\,MeV) and because of that the chiral symmetry is only approximate. Nonetheless, the EFT that describes (massless) pions is quite useful. It is called {\em chiral perturbation theory} and is an example of a NLSM. 
 
 \vspace{0.8mm}
In our presentation of the amplitude approach to bootstrap the landscape of QFTs in Sections \ref{s:ampboot} and  \ref{s:ampbootex}, we focus on pion NLSMs. 
\end{itemize}

We have given a few of examples of QFTs, but this is far from the whole picture.  In the examples, I emphasized in each case the dependence on energy scale. The way these theories behave with change in scale is governed by {\em renormalization group (RG) flow}. RG flow is a way to move around in theory space; some theories are connected with RG flows, others are not. We have seen that some theories (like QED) become trivial at low energies, while others (like QCD) become strongly coupled. There is also an option that some theories flow to non-trivial fixed-points at low-energies. This for example is the case for QCD with a sufficiently high number of families of quarks. At an RG fixed point, the physics no longer changes with scale.  In many cases, scale invariance is associated with enhanced spacetime symmetry, namely {\em conformal symmetry}. 

\subsection{Conformal Field Theories}
\label{s:CFT}
Conformal Field Theories (CFTs) are characterized by having, in addition to   Poincar\'e symmetry, conformal boost symmetry. One way to describe it is that special conformal boosts are transformations that preserve angles. Another way is to consider the inversion transformation that sends a spacetime coordinate $x^\mu$ to $x^\mu/x^2$, where $x^2=-t^2+|\vec{x}|^2$ is the relativistic invariant spacetime distance. (We use conventions of $c=1$ throughout.) Then a special conformal boost is what you get when you do inversion, followed by a translation, followed by another inversion.

Sounds a little complicated? OK, let us try to get a little intuition. 
Inversion sends $x^2$ to $1/x^2$. Ignoring the time-component of $x^\mu$ we then see that inversion trades long distance with short distance, and vice versa. In a relativistic theory, this then interchanges low energy and high energy. For a theory to have such a property,\footnote{CFTs need not have inversion symmetry, for example chiral CFTs do not, but for the purpose of this introduction let us just consider CFTs that are invariant under inversion.} it cannot have any preferred  scales because inversion symmetry requires that the physics above and below a given scale has to be the same. For instance, if a particle has mass $m$, then for energies  $E \ge 2 m c^2$ such particles can be pair-produced, but for $E < 2 m c^2$ they cannot. Thus  physics is different above and below $2mc^2$ and hence masses cannot be allowed with inversion symmetry. Similarly no other dimensionful parameters are allowed. Hence, a relativistic theory with inversion symmetry is necessarily also {\em scale invariant}. This is the aspect of CFTs that is most relevant for us in this presentation: CFTs are scale invariant.\footnote{The reverse may in general not be true: scale invariance does not in general imply conformal invariance.}

Scale invariance is very different from our everyday experience. It is not the same to take a ice bath as it is to put your finger in boiling water (and neither is pleasant for very long). People age differently over the time scale of a year than they do over ten years. 
So it may seem like conformal symmetry, or even just scale invariance, is a property very different from what we encounter in our everyday world. That is true, nonetheless, scale invariance does happen in nature --- and it can be found it in the lab. 

We already encountered an example of scale invariance in Section \ref{s:watermagnet}. Recall that near the critical point of water (or the Curie point for the ferromagnet), the correlation length $\xi$ diverges so that all length scales become of equal relevance and the system becomes scale invariant at the critical point. 
There is no proof that it also becomes conformal, however, in the Section \ref{s:introbootstrap} we describe how modeling the critical point using conformal field theory techniques (specifically here for the {3d Ising model}) makes it possible to determine critical exponents such as $\nu$ in \reef{xi}. 

Other examples of CFTs arise in 4-dimensional supersymmetric gauge theories. 
In Section \ref{s:exQFTs}, we noted that gluons, the spin-1 massless particles of the strong nuclear force, are described by Yang-Mills theory. 
Yang-Mills theory is not a conformal theory, but the gluons can be coupled to other particles in such a way that the resulting theory is conformal.  A useful tool in this context is supersymmetry, a symmetry that partners bosons and fermions into supermultiplets in which all particles must have the same mass and their interactions are quite restricted. The massless fermion superpartners of gluons are called gluinos and the QFTs describing them are called {\em super Yang-Mills (SYM) theories}. In models without gravity, the gluons can have either $\mathcal{N}=0,1,2$ or $4$ gluino partners.\footnote{\label{footieN3} Why not $\mathcal{N}=3$, you ask? Sure, $\mathcal{N}=3$ is fine too. For a Lagrangian theory, CPT invariance (charge conjugation, parity, time-reversal) implies that $\mathcal{N}=3$ SYM theory is equivalent to $\mathcal{N}=4$ SYM.}
For the $\mathcal{N}=2$ or $4$ cases, there also needs to be massless scalars coupled supersymmetrically to the gluons and gluinos. 

The massless $\mathcal{N}=4$ SYM theory in 4d is very special: its Lagrangian is completely fixed by supersymmetry and the couplings do not run at all with scale. Not only is 
{\em $\mathcal{N}=4$ super Yang-Mills theory} scale-invariant, it is in fact also conformally invariant, even at the quantum level. It is a theory that in many respects is considered ``the simplest" QFT due to its strong constraints from symmetries that allows calculational control that one often lacks in other theories. 
$\mathcal{N}=4$ super Yang-Mills theory appears in many areas of theoretical high energy physics, often as a ``testing lab" for developing new techniques and gaining insights into similar systems, for example QCD.

There is a multitude of other known CFTs. Coupling $\mathcal{N}=2$ SYM supersymmetrically to matter multiplets can give rise to superconformal (supersymmetric and conformal) theories (SCFT). The subject of  $\mathcal{N}=2$ SCFTs is in itself very rich, with some $\mathcal{N}=2$ SCFTs described by Lagrangians and others having only strongly coupled dynamics and no Lagrangians. Understanding the landscape of $\mathcal{N}=2$ SCFTs is in itself an actively investigated research subject. 

There is a multitude of other known CFTs and SCFTs. Some arise in condensed matter systems, others in string theory. Recall that (S)CFTs are the light beacons in the much vaster landscape of QFTs, so it is noteworthy that they  themselves make up a rich landscape that is even far from fully explored. 

\subsection{QFT Observables}
The key observables in QFTs are {\em correlation functions}. Correlation functions are familiar from cosmology where they measure the correlations between different locations in the map of the cosmic microwave background. They are familiar from condensed matter physics where we may be interested in whether interactions are short distance or long distance; in fact the correlation length $\xi$ we discussed in Section \ref{s:watermagnet} is the characteristic length associated with a 2-point (connected) correlation function 
\be
  \< \sigma(x) \sigma(y) \> \sim e^{-|x-y|/\xi}\,
\ee  
for $|x-y| \gg \xi$. It measures the correlations between interactions between locations $x$ and $y$. In a lattice model, the field $\sigma(x)$ can be thought of as designating whether the site at position $x$ has spin up or down (say $\sigma(x) = \pm 1$). 
In general, the $n$-point correlation function measures the correlation between quantities  at $n$ different spacetime locations.  

One approach to correlation functions is to reduce them from their general form to an ``on-shell" form in momentum space; this gives the on-shell scattering amplitudes from which one can compute the observable scattering cross-sections. The amplitudes are the observables of interest in the Scattering Amplitudes Program. Another approach to study correlation functions to impose on them symmetries and mathematical consistency conditions: that is the what is done in the Conformal Bootstrap Program in the context of CFTs.

\section{Introduction to the Modern Scattering Amplitudes Program}
\label{s:amps}

A scattering experiment consists of banging things together to see what comes out. For example at the LHC protons are collided against protons at about 10,000 times their rest mass. At the microscopic level, the partons (gluons and quarks) inside the protons interact with each other in the collisions. A  process representative for the physics we discuss here is the process of two gluons interacting to produce a new set of gluons, e.g.~
\be
   \label{6gluons}
    g + g ~\to~ g + g + g + g 
\ee
The gluons are described by Yang-Mills theory. At sufficiently high energies (such at the LHC) it is weakly coupled and it therefore makes sense to study the scattering of gluons (and other partons) perturbatively. Eventually the partons hadronize and form jets of mesons and baryons. Here we focus on the high-energy part of the process that just involves gluons scattering inelastically to gluons.

The probability for a scattering process to occur is encoded in the scattering amplitude $A_n(i \to f)$, where $n$ is the total number of initial and final state particles in the process $i \to f$. It is related to the experimentally measurable observable, the differential scattering cross-section as
\be
  \frac{d\sigma}{d\Omega} \sim \int |A_n(i \to f)|^2\,.
\ee
The integral here is taken suitably over phase space. For a given initial state $i$, $d\sigma/d\Omega$ gives the probability of measuring the final state $f$ as a function of scattering angles and energies. 

Amplitudes are traditionally calculated as the sum of Feynman diagrams constructed from the vertices and propagators of a theory described by some Lagrangian. The perturbative expansion in small couplings is organized diagrammatically in a loop-expansion where the leading order is tree-level, the first correction is the sum of one-loop diagrams, the next correction consists of the two-loop diagrams etc. A general rule of thumb: the more loops and the more particles, the harder it is to calculate the amplitudes. For more particles, this is because the number of diagrams tends to grow combinatorially. For higher loops, the number of diagrams is a significant issue too, but not the only one; the evaluation of highly non-trivial integrals over energy-momentum running in the closed loops can be a challenging roadblock as well. 

Amplitudes depend on {\em ``external data"}: the momenta $p_i^\mu$ for each of the external particles and polarization vectors for spin 1 particles and spin wavefunctions for fermions. 
The external momenta $p_i^\mu$ are subject to the requirements of being on-shell and satisfying momentum conservation, i.e.~($c=1$)
\be
\label{genmomcons}
  p_i^2 \equiv -E_i^2 + |\vec{p}_i|^2 =  m_i^2
  ~~~~\text{and}~~~~ \sum_{i\in \text{incoming}} p_i^\mu = \sum_{i\in \text{outgoing}} p_i^\mu\,.
\ee
Amplitudes with \reef{genmomcons} satisfied and polarizations or wavefunctions are included as appropriately for all external particles are called 
{\em on-shell amplitudes}.

To calculate the process \reef{6gluons} at the leading order in perturbation theory, one needs to add up all the tree Feynman diagrams with 6 external gluons built from the Feynman rules extracted from the Yang-Mills Lagrangian. There are cubic and quartic gluon self-interactions so one gets 
\be
   A_6 =    
      \raisebox{-0.55cm}{
    \begin{tikzpicture}[scale=0.3, line width=1 pt]
    	\draw [gluon] (-2,2)--(0,0);
	\draw [gluon] (-2,-2)--(0,0);
	\draw [gluon] (0,0)--(4,0);
	\draw [gluon] (4,2)--(4,0);
	\draw [gluon] (4,-2)--(4,0);
	\draw [gluon] (9,2)--(8,0);
	\draw [gluon] (9,-2)--(8,0);
	\draw [gluon] (8,0)--(4,0);
    \end{tikzpicture}
    }
   ~+ 219~\text{more diagrams}.
\ee
For scattering of 7 gluons one needs 2485 diagrams and for 8 gluons it would be 34,300 diagrams. 
Each diagram translates into a rather complicated mathematical expression via the Feynman rules. Calculating these amplitudes by hand using Feynman diagrams is not a great way to spend your time. And this is still only the leading order in perturbation theory, imagine the complications at loop-level!

 One of the goals of the modern on-shell amplitudes program is indeed to come up with better and more efficient ways to calculate scattering amplitudes, but there are also several other avenues of progress. 
 
\subsection{Modern Amplitudes Program}

Early modern approaches to scattering amplitudes were pioneered by Bern, Dixon, Kosower, often  driven by applications in particle phenomenology. This thrusts continues as the fields has broaden significantly in the past $\sim17$ years and has attracted many more people to the field. We outline five directions in modern research on scattering amplitudes:
\begin{enumerate}
\item {\bf New Computational Techniques.} 
One goal of the amplitudes program is to develop new calculational techniques to facilitate more efficient computation of scattering amplitudes --- and to perform calculations of amplitudes that may even be impossible using traditional Feynman diagrammatics. At tree-level, examples of such new techniques are {\em on-shell recursion relations} (they go under names such as BCFW \cite{Britto:2004ap,Britto:2005fq}, CSW expansion \cite{Cachazo:2004kj}, all-line shifts \cite{Cohen:2010mi,Cheung:2015cba}, soft shift recursion \cite{Cheung:2015ota,Elvang:2018dco} etc). Some on-shell recursions are quite general and others are more closely adapted to the field theory they are applied to. The general idea is to recycle lower-point amplitudes into higher-point ones. For example, 3-particle scattering determines 4-particle scattering. Then 3- and 4-particle amplitudes can be recycled into 5-particle scattering etc. Sometimes the recursion relations can be solved exactly, there are for example closed-form expressions for scattering of any $n$ number of gluons at tree-level \cite{Drummond:2008cr}. 

\vspace{0.8mm}
The derivation of on-shell recursion relations exploits knowledge of the analytic structure of amplitudes. Tree amplitudes are rational functions of the external data and they have simple poles where physical particles can be exchanged. On these poles, unitarity guarantees that the tree amplitudes factorize into products of lower-point amplitudes. The information of the location of the poles in momentum space and the factorized form of their residues are the basis of the   on-shell recursion relations. 

\vspace{0.8mm}
Loop-amplitudes have a more complicated analytical structure. While they can have rational terms, they generally also involve more complicated analytical functions such as logarithms, polylogarithms, and worse. One powerful technique is the method of  {\em generalized unitarity} (pioneered in \cite{Bern:1994zx} and in since applied in countless contexts; see the review \cite{Bern:2011qt} and references therein). Here one uses that the integrand of a loop amplitude is a rational function that develops poles for specific choices of the loop-momenta. On these poles, the integrand factorizes into amplitudes with fewer loops. At 1-loop level, such ``cuts" (and their $d$-dimensional generalizations) can be used to determine the integrand from tree-level amplitudes. At 2-loop level, the 1-loop and trees are recycled into determining the cuts, and so on. 

\vspace{0.8mm}
The fact that the loop-integrand is rational can  be used to construct on-shell recursion relations at loop-level in certain models, such as for the planar limit of $\mathcal{N}=4$ SYM \cite{ArkaniHamed:2010kv}. 
\item {\bf Mathematical Structure and Geometry.} We have already alluded to how the mathematical structure of amplitudes inform the development of calculational tools. But on-shell amplitudes themselves also harbor hidden structures that cannot be inferred from the Lagrangian. For example, many amplitudes have analytic expressions that are much simpler than the Feynman diagram representations would suggest. What is responsible for such simplifications? Elucidating the mathematical structure of amplitudes, uncovering hidden symmetries, reformulating the scattering problem in novel mathematical terms is another goal of the amplitudes program. 

\vspace{0.8mm}
Recent ideas include representations of amplitudes in terms of contour integrals in Grassmannian spaces (which are spaces of $k$-planes in $n$-dimensional space)  \cite{ArkaniHamed:2009dn} and geometrizations such as polytopes \cite{Hodges:2009hk,ArkaniHamed:2010gg}, amplituhedrons \cite{Arkani-Hamed:2013jha},  associahedrons and more generally positive geometry \cite{Arkani-Hamed:2017mur}. The idea is that the amplitude is related to a volume form for a  geometric object in some abstract mathematical space.  The boundaries of the geometric object correspond to the location of poles. Different triangulations of the volume of this object can be mapped to different equivalent mathematical formulas for the amplitudes, one is the Feynman diagram representation, some correspond to the results of on-shell recursion relations, and   others again are inherently different. This is explored at both tree- and loop-level. 

\vspace{0.8mm}
Through the connection to interesting mathematical structures, there are now fruitful collaborations between the amplitude community and mathematicians on subjects such as  positive geometry and cluster algebras. 
\item {\bf Exploring the Space of QFTs: Amplitude Bootstrap.} Traditionally one starts with a Lagrangian, writes down the Feynman rules, and use them to compute the amplitudes. Any symmetries of the Lagrangian manifest themselves on the amplitudes as ``Ward identities".
 Here is an example: if the Lagrangian has a symmetry that gives charge conservation, the associated Ward identity says that the amplitude of any process that violates charge conservation has to vanish. 

\vspace{0.8mm}
A new approach to QFTs is to turn this logic on its head and instead of starting with the Lagrangian, one takes the physical observables, the amplitudes in particular, as the starting point, impose constraints on particle spectrum and symmetries on the amplitudes and subject them to tests of mathematical consistency. This then allows one to explore the existence of QFTs with the assumed properties. In particular, it gives a systematic way to explore the landscape of possible theories with a set of specified symmetries.  We describe this further in Section \ref{s:ampboot} and in more detail in the technical Section \ref{s:ampbootex}.

\item {\bf Double-Copy.} In the mid-80s it was realized that tree-level closed string amplitudes can be written as sums of products of tree-level open-string amplitudes \cite{Kawai:1985xq} --- these are known as the Kawai-Lewellen-Tye (KLT) relations. In the limit of infinite string tension, this becomes the field theory statement that the graviton tree amplitudes can be obtained as a sum of products of gluon scattering amplitudes. This relation is sometimes written 
\be
  \text{``gravity~~=~~(gauge theory)$^2$\,"} 
\ee
and is by now referred to as an example of the {\em double copy}.

\vspace{0.8mm}
Starting in 2008, it became clear that there is more to this story. Bern, Carrasco, and Johansson \cite{Bern:2008qj} found that tree-level gauge  theory amplitudes of gluon scattering could be written in a form where certain kinematic numerators obey the same Jacobi identities as the algebraic color-factors of the non-abelian gauge group of the theory. This is called {\em color-kinematics duality}. Moreover, if one replaces the color factors in this representation of the amplitude with the kinematic factors of  gauge theory, remarkably the result is the gravity tree amplitude! This is the {\em  BCJ double copy} and it has since been generalized to amplitudes of other field theories (e.g.~\cite{Cachazo:2014xea}). 
BCJ conjectured (and it has been tested in multiple contexts) that a similar color-kinematics  prescription and double-copy also holds at the level of the loop-integrand \cite{Bern:2010ue}. 

\vspace{0.8mm}
Since tree-level scattering represents the classical physics, it is natural to explore if there is a similar way to double copy solutions to the classical equations of motion. For example, the double-copy of a Coulomb-type solution in gauge theory to a black hole solution in General Relativity as a weak-field expansion. This direction has thus attracted attention of researchers from other  fields, such as theorists studying classical solutions in General Relativity and supergravity and cosmologists.  For a  recent review of the double-copy and its applications, see \cite{Bern:2019prr}.

\item {\bf Gravitational Wave Physics.} With the recent detection of gravitational waves from black hole inspirals made by the LIGO detector, and the 2017 Nobel Prize to the pioneers of experimental gravitational wave physics, the field of gravitational waves has received intense interest. Remarkably, amplitude techniques prove very useful here too. Starting with the Einstein equation, it is not hard to derive the linearized solution for a freely propagating gravitational wave. It is, however, a highly non-trivial matter to model the gravitational wave radiation resulting from the inspiral of two heavy objects such as black holes or neutron stars. Numerical breakthroughs have been an essential part of the study.  There are also many other techniques, such as effective field theory formulations \cite{Goldberger:2004jt}. 

\vspace{0.8mm}
On the analytic side, one uses a post-Minkowskian or post-Newtonian expansion for a Hamiltonian with an effective potential; the corrections here are in terms of powers of Newton's constant and orbital speed $v/c$, respectively. The inspiral problem is an elliptical solution to this Hamiltonian: they are the bound states. At first sight this has absolutely nothing to do with scattering amplitudes. However, there are also hyperbolic solutions: a classical example of the hyperbolic problem is the famous deflection of light by a heavy object. 

\vspace{0.8mm}
The deflection (i.e.~hyperbolic) problem is basically a scattering process: two massive objects are in the initial state, they interact and then fly apart again after some exchange of energy-momentum. We can compute the scattering of massive particles under exchange of gravitons. The higher order graphs in such a calculation are in direct correspondence with the higher-order corrections in the effective Hamiltonian, so the effective Hamiltonian can be reconstructed from the scattering amplitudes and then used to study the bound state problem. If this were done with Feynman rules there would be limited calculational advantage. However, with the modern on-shell amplitude machinery, very promising progress has been made. At this stage, an example in this direction includes the calculation of the Hamiltonian for massive spinless binary systems to 3rd post-Minkowskian order (meaning order $G^3$ in the Newton coupling) \cite{Bern:2019nnu,Bern:2019crd,Bern:2020buy}. There are also approaches \cite{Kalin:2019rwq,Kalin:2019inp} that try to circumvent  the construction of the effective potential and directly get gravitational wave information from the scattering amplitudes as well as related EFT approaches \cite{Kalin:2020fhe}. This is a rapidly developing field that has facilitated fruitful interactions between the community of   General Relativity theorists and the amplitude community. 

\end{enumerate}

There are many other very interesting developments in the field of scattering amplitudes. One is the system of {\em scattering equations} that has led to the so-called CHY construction of amplitudes that comes with its own formulation of the double-copy and has led to the realization of new examples of its application \cite{Cachazo:2013hca,Cachazo:2013iea,Cachazo:2014xea}. Another direction has been the connection between the universal {\em soft behavior} of  gravitons, and the infinite-dimensional BMS symmetry in General Relativity \cite{He:2014laa}. There are other types of amplitude bootstraps too, such as the {\em integrability} approach \cite{Basso:2013vsa}, the {\em loop-amplitude bootstrap} using cluster algebraic structures \cite{Caron-Huot:2020bkp}, and the {\em S-matrix bootstrap} that exploits the conformal bootstrap \cite{Paulos:2016fap}. The latter is an example of the fruitful overlap between the conformal bootstrap and amplitudes communities. Furthermore, phenomenologists are increasingly using amplitude techniques to study physics beyond the Standard Model, for example to organize higher-derivative operators in {\em Standard Model Effective Field Theory (SMEFT)} \cite{Shadmi:2018xan,Craig:2019wmo}.  Finally, using basic properties of amplitudes, one can prove a number of interesting general theorems about QFTs such as \cite{Benincasa:2007xk,McGady:2013sga,Elvang:2016qvq,Elvang:2013cua,Elvang:2015rqa}: 
\begin{itemize}
\item There can be no theories in flat space with massless particles of spin greater than 2 interacting with gravity. 
\item There can only be one graviton-field (i.e.~only one massless spin-2 particle), it must self-interact and it must couple exactly the same way to any other particle (the equivalence principle).
\item A spin 3/2 particle must couple supersymmetric to the spin-2 graviton.
\item Spin-1 massless fields can only self-interact if there is a Lie algebra structure with 3-index fully antisymmetric structure constants, and
\item A $\mathcal{N}$=8 superconformal 3d theory requires the existence of fully 4-index antisymmetric structure constants \cite{Huang:2010rn}.
\end{itemize}

As this hopefully illustrates, the field of amplitudes concerns a diverse range of subjects. At this point, the annual Amplitudes conference, now in its 12th year, attracts a few hundred international participants (and even more in its recent online Zoomplitudes version). We have highlighted here some general directions and current areas of interests, but of course this is in no way complete. The interested reader may want to consult the several newer reviews and textbooks on modern methods in amplitudes, see for example \cite{Elvang:2013cua,Elvang:2015rqa,Henn:2014yza,Dixon:2013uaa,Cheung:2017pzi}.

I am now switching gears to discuss in a little more detail one direction that shares ideology with the conformal bootstrap program. In the following, I  outline the ideas, then in Section \ref{s:ampbootex} I provide a very detailed example of its implementation and results. 

\subsection{Amplitude Bootstrap on the Space of QFTs}
\label{s:ampboot}

Suppose someone asks:
\begin{quote}
 {\em ``Does there exist a local relativistic QFT with two massless real scalars such that every tree amplitude vanishes in the limit where a single momentum is taken soft?  Is such a model unique? Must it have any particular symmetries, such as an interchange symmetry that requires the scalar particles to be on equal footing?".} 
\end{quote} %
The vanishing soft limit means that $A_n \to 0$ as $p_i^\mu \to 0$ for any on-shell external momentum $i=1,2,\ldots,n$. These soft limits are called Adler zeros and were first discovered in the context of pion scattering \cite{Adler:1964um}. 

The physical context of this question is that such a model describes the low-energy dynamics of two massless Goldstone scalar particles arising from some spontaneous symmetry breaking. For any explicitly given symmetry breaking patterns of some symmetry group $G$ to a subgroup $H$, there are techniques for systematic construction of Lagrangians of the Goldstone modes. But for more open-ended questions aimed at understanding the space of possible theories and any additional emergent symmetries they may have, the Lagrangian approach is limited and often complicated by field-redefinitions that can obscure symmetry properties. 

A traditional `bottom-up' approach to such a question is to try to write down a Lagrangian with kinetic terms for the two scalar fields and some local interactions that preserve the desired symmetries. Suppose we fail to construct a Lagrangian with these properties: does it mean that such a theory does not exist? Or did we miss out on some smart way to do this? 
Or suppose we did succeed in writing down a Lagrangian with these properties, is it unique? Or did we miss some other allowed interactions? Are there different ways to write the Lagrangian that nonetheless result in exactly the same observables? For example by being related by field redefinitions.

A modern approach to such questions is to start with the physical observables, namely the amplitudes. A clear advantage of this approach is that the amplitudes are independent of field redefinitions (and in case of gauge fields, the amplitudes are gauge invariant so gauge-choices and so on do not matter). The symmetries of the model manifest themselves on the amplitudes via  Ward identities; linear relationships among amplitudes, valid either at all generic momenta or in certain momentum limits.

In the on-shell approach, one imposes on the amplitudes a set of assumptions about the particle spectrum of the model and its symmetries (exact or spontaneously broken) as well as mathematical consistency on the amplitudes. At tree-level, consistency conditions refer to properties like:

\begin{tabular}{lcp{11cm}}
{locality} ~~&$\implies$&{correct simple poles corresponding to exchanges of physical particles; no spurious (unphysical) poles.}
\\
{unitarity} ~~&$\implies$&{the residues on the simple poles factorize into lower-point on-shell amplitudes.}
\end{tabular}

One starts with the most general ansatz for the lowest-point amplitudes subject to the symmetries. As the lowest-point amplitudes in the model, they cannot have any physical poles since there is no lower-point amplitudes they can factorize into. So they must be polynomials in the external data (momenta, polarizations etc) and each independent polynomial  corresponds to an independent interaction term in an associated Lagrangian. Independence means under the use of momentum conservation and other algebraic identities; at the level of the Lagrangian, momentum conservation simply translates to integration-by-parts. 

Next fuse the lowest-point amplitudes together to make higher-point amplitudes, for example via a recursion relation of some valid form. The higher-point amplitude must have the required symmetries too. This may fix constants in the parametrization of the amplitudes. It may even set the amplitudes to zero. If all constants are set to zero by the mathematical consistency conditions, it means that there are no amplitudes that respect the requested symmetries and hence there can be no such non-trivial QFT. For if there were, it would produce non-vanishing scattering amplitudes. On the other hand, if all imposed mathematical consistency checks are satisfied with some non-vanishing scattering amplitudes, then it is evidence that {\em perhaps} such a QFT may exist. It cannot be a definitive ``yes" because there could be further restrictions arising at higher-points and one would have to work harder to {\em prove} existence of a field theory. 

What we have described here is the ``amplitude bootstrap". It is very powerful as a tool to rule out the existence of theories with too strong symmetry requirements, but cannot say ``yes, it does exist" without further input. As it turns out, this ability to answer ``no" or ``maybe" is one thing it has in common with the conformal bootstrap, as we shall see in Section \ref{s:introbootstrap}.

Let us illustrate the idea briefly using a combination of Lagrangian reasoning and on-shell amplitudes. We want a model of two massless scalars $\phi_1$ and $\phi_2$  such that the amplitudes vanish in the limit where any one of the particle momenta goes soft, i.e.~$p^\mu_i \to 0$ for any one of the external momenta in the process. Any model with $\phi^4$-type interaction would fail the criteria since the 4-point amplitude is a constant and therefore does not vanish for any choice of momentum. What about an interaction term like 
$\phi_1^2 (\partial \phi_2)^2$? Since $\partial_\mu \to i p_\mu$ and $p_i^2 = 0$, this gives a 4-point amplitude 
$A_4(\phi_1 \phi_1 \phi_2 \phi_2) = 2 p_3 \cdot p_4 = (p_3 + p_4)^2$ (the momenta are labeled 1,2,3,4 and related to the particles as the order in which they are given in the amplitude). This vanishes for any one of the momenta going to zero since the massless particles have $p_i^2=0$ and momentum conservation ensures $(p_1 + p_2)^2 = (p_3 + p_4)^2$. 

So we are good, right? Not so fast! At 6-point, the amplitude $A_6(\phi_1\phi_1\phi_1\phi_1\phi_2\phi_2)$ includes diagrams like
\be
  \label{A6diagex}
     \raisebox{-0.58cm}{
    \begin{tikzpicture}[scale=0.4, line width=1 pt]
    	\draw (-1,1)--(0,0);
	\draw (-1,-1)--(0,0);
	\draw[dashed] (-1.5,0)--(3.5,0);
	\draw  (2,0)--(3,1);
	\draw (2,0)--(3,-1);
	\node at (-1.4,1) {\small $1$};
	\node at (-1.9,0) {\small $5$};
	\node at (-1.4,-1) {\small$2$};
	\node at (3.5,-1) {\small$4$};
	\node at (3.9,0) {\small $6$};
	\node  at (3.5,1) {\small $3$};
	\node at (1.17,0.55) {\small $P$};
    \end{tikzpicture}
    }
    =(p_1 + p_2)^2\frac{1}{(p_1 + p_2 + p_5)^2}(p_3 + p_4)^2\,,
\ee
where the solid lines indicates the $\phi_1$-particles and the dashed line the $\phi_2$-particles. In the limit where $p_5^\mu \to 0$, the diagram gives  $(p_3 + p_4)^2 = 2 p_3\cdot p_4$ which is non-zero for generic momenta. So even though we had a 4-particle interaction that did the job we wanted for 4-point amplitudes, it failed at 6-point. Now there are of course other diagrams that contribute too: for example, we can exchange $(2 \lra 3)$ and $(2 \lra 4)$ in \reef{A6diagex}. Together with \reef{A6diagex}, these then contribute 
\be
  \label{interm1}
   2 p_3\cdot p_4 + 2 p_2\cdot p_4 + 2 p_3\cdot p_2
   = (p_2 + p_3 + p_4)^2
\ee
to the $p_5 \to 0$ soft limit of the amplitude. But there are also diagrams where line 1 is not on the same vertex as the soft line 5. They contribute 
\be
  \label{interm2}
   2 p_1\cdot p_2 + 2 p_1\cdot p_3 + 2 p_1\cdot p_4
   = 2 p_1 (p_2 + p_3 + p_4).
\ee
Thus the sum of the pole diagrams contribute the sum of \reef{interm1} and \reef{interm2}
\be
  (p_1+ p_2 + p_3 + p_4)^2 = (p_5 + p_6)^2
\ee
by momentum conservation. This is generically non-zero, so there is no way to get zero if only these 4-point interactions are included. We need 6-particle interactions in order to cancel the non-vanishing results from diagrams such as \reef{A6diagex}. This can for example be engineered from a Lagrangian interaction term of the form $\phi_1^4(\partial \phi_2)^2$ whose coefficient is tuned such that its contribution to the 6-point amplitudes exactly cancels that of the pole terms like \reef{A6diagex} in the soft limit. Along with other terms needed at 6-point order, the continued consistency with vanishing soft limits at higher points ends up fully dictating the model!

As for being on equal footing, it would appear from the above construction that $\phi_1$ and $\phi_2$ are not interchangeble. However, when one is more careful about setting up the problem, a symmetry between $\phi_1$ and $\phi_2$ does in fact emerge. This is shown in full detail in Section \ref{s:ampbootex}.

For now, what I wanted to illustrate here was the logic of how the constraints are imposed on the amplitudes and how it allows us to learn about the structure of the model. In the more general approach, we parameterize all 4-particle amplitudes in the most general way and then impose the soft constraints on the 4- and higher-point amplitudes (Section \ref{s:ampbootex}). This allow us to get ``no, does not exist" or ``maybe" as the answer to whether such a QFT exists. This is very similar to how the conformal bootstrap works, as we now explain.

\section{Introduction to the Modern Conformal Bootstrap}
\label{s:introbootstrap}

To set up a  scattering problem,  we start at time $t = - \infty$ with initial state particles that are infinitely far apart and non-interacting (i.e.~free). Then the particles come together, scatter through some (weak) interaction process, and end up far apart as free particles in the far future, $t= +\infty$. These so-called initial and final asymptotic states of the far past and future are the in and out states of the perturbative scattering amplitudes. 

In a conformal field theory, there is no scale, so there is no sense of ``far apart" or ``far into the past/future". A CFT is scale invariant and has no asymptotic states. For that reason, a scattering amplitude is a priori not a good observable.\footnote{Nonetheless, amplitudes in some CFTs, such as in $\mathcal{N}=4$ SYM, play a central role in the amplitudes program. They can be understood as limits of amplitudes in a a non-conformal QFT.} Moreover, if we wish to ask ``what CFTs exist?" we have to include also theories without weakly coupled limits, i.e.~those in which we cannot talk about the ``free" particle states because the interactions are always strong. 

The {\em conformal bootstrap} is a method to explore the landscape of CFTs without need for Lagrangians, weak couplings, or the concept of particles or scattering amplitudes; instead physical and mathematical consistency constraints are imposed on the observables in the CFT, namely the correlation functions of local operators. 

\subsection{Observables and CFT Data}

Correlation functions are the central observables in a CFT. So, what does that mean? Classical fields are familiar from 4d electromagnetism: at each point in space and time, the electric $\vec{E}(\vec{x},t)$ and magnetic $\vec{B}(\vec{x},t)$ fields give a vector-valued result for the strength and direction of the electromagnetic field. They are local fields in that they depend on a single point in spacetime. Under Lorentz transformations, the electric and magnetic fields are mixed, and in relativistic contexts it is useful to combine them into the field strength $F_{\mu\nu}$, where in a given inertial frame the components of the electric and magnetic fields are $E_i = F_{ti}$ and $B_i = \sum_{j,k=1}^3 \eps_{ijk} F_{jk}$, Here $\eps_{ijk}$ is fully antisymmetric in its indices and  $\eps_{123}=1$. The field $F_{\mu\nu}$ is an example of an operator with non-zero spin. We can form other operators from the field strength, such as  $F^2 =  F_{\mu\nu} F^{\mu\nu}$ and $F^4$ etc. Examples of other operators are spin-0 scalar fields $\phi(x)$ and spin-1/2 fermion fields $\psi(x)$ and powers thereof, i.e.~$\phi^n$ and $\psi^2$. We can take derivatives of operators to get new ``descendant" operators, such as $\partial_\mu \phi = \partial \phi/\partial x^\mu$ etc. 
These are examples of operators based on local fields, but more generally we consider local operators in a completely abstract sense. We denote such operators as $\mathcal{O}_i$, where $i$ is a collective index that includes both operator type and any Lorentz indices it may have. 

A property of operators that is important for the our discussion is their {\em scaling dimension}. Under a scaling $x^\mu \to \lambda x^\mu$ of the spacetime coordinates, an operator scales homogeneously as 
\be
  \label{defDelta} 
  \mathcal{O}_i(x) \to \lambda^{\Delta_i} \mathcal{O}_i(\lambda x)\,.
\ee
 Consider a scalar field $\phi$ in a $d$-dimensional  free theory. The action just has the kinetic term
\be
  S = \int d^d x\, \partial_\mu \phi \partial^\mu \phi \,.
\ee
The scalar field has mass-dimension $[\phi]=(d-2)/2$ in order for the action to be dimensionless ($\hbar=c=1$). Performing a trivial change in integration variable $x \to x' = \lambda x$, we see that we must also have $\Delta_\phi = (d-2)/2$.
As in this example, the scaling dimension $\Delta_i$ is the same as the mass-dimension of an operator $\mathcal{O}_i$ in a free theory. However, in an interacting quantum theory, $\Delta_i$ receives quantum corrections and generally departs from the free value. Hence $\Delta_i$ is considered a real number in the following. 

In a conformal field theory, the symmetries are so powerful that the 1-, 2-, and 3-point correlation functions are completely fixed up to a set of constants in the 3-point correlator. 1-point functions vanish $\<\mathcal{O}_i(x)  \> = 0$. 
We write the 2-point and 3-point correlators of operators $\mathcal{O}_i$ diagrammatically as 
\be
 \big \< \mathcal{O}_i(x) \mathcal{O}_j(y) \big\> = 
     \raisebox{-0.2cm}{
    \begin{tikzpicture}[scale=0.4, line width=1 pt]
	\draw (-1,0)--(1.2,0);
	\node at (-1.9,0) {\small $i$};
	\node at (2.3,-.1) {\small $j$};
    \end{tikzpicture}
    }
    ~~~~
    \text{and}
    ~~~~~
     \big \< \mathcal{O}_i(x) \mathcal{O}_j(y) \mathcal{O}_k(z) \big\> = 
         \raisebox{-0.6cm}{
    \begin{tikzpicture}[scale=0.4, line width=1 pt]
    	\draw (-1,1)--(0,0);
	\draw (-1,-1)--(0,0);
	\draw (-0,0)--(1.2,0);
	\node at (-1.4,1) {\small $i$};
	\node at (-1.4,-1) {\small$j$};
	\node at (1.6,0) {\small $k$};
    \end{tikzpicture}
    }
\ee
Operators with different scaling dimensions are orthogonal in the sense that their 2-point functions vanish. If there are degenerate operators with the same operator dimension, they can be organized in a basis such that their 2-point correlators vanish for distinct operators: 
\be
\big \< \mathcal{O}_i(x) \mathcal{O}_j(y) \big\> = 0 ~~~
\text{for}~~~i \ne j \,.
\ee

The unfixed constant in the 3-point correlator is called the ``OPE coefficient". This comes from the idea of the Operator Product Expansion (OPE) which states that in a local theory, the fluctuation generated by a product of operators in close proximity should be expressible as a sum of local operators. The coefficients of the operators in that sum are the OPE coefficients $c_{ijk}$: so as $y \to x$
\be
  \label{simpleOPE}
  \mathcal{O}_i(x) \mathcal{O}_j(y) \sim \sum_{n}  c_{ijn} \,\mathcal{O}_n(x) \,.
\ee
(We defer discussion of some of the spacetime dependence to Section \ref{s:confboot}.)
The subscripts $i,j,n,\dots$ is some collective index that labels the operators in the CFT.
The sum on the right is in principle over all operators in the theory, but some coefficients may be zero. For example, a fermionic operator would not appear in the OPE of two bosonic operators.

Multiplying \reef{simpleOPE} by $\mathcal{O}_k(x)$ and taking the expectation value (think of the analogue to undergraduate quantum mechanics here) selects on the RHS the coefficient $c_{ijk}$ via the ``orthogonality" property of the 2-point correlation function,
$\big \< \mathcal{O}_k(x) \mathcal{O}_n(y) \big\> \sim \delta_{kn} \times$(function of $|x-y|$).
 Hence one finds that the LHS 
 $\big \< \mathcal{O}_i(x) \mathcal{O}_j(y) \mathcal{O}_k(z) \big\>$ is  determined by the OPE coefficient $c_{ijk}$. In the above, we are completely glossing over the details of the dependence of the spacetime coordinates $x,y,z$;  those who wish to see a more detailed account can find it in the technical review in Section \ref{s:confboot}.

In order to explore the landscape of CFTs, we have to describe what characterizes a CFT. For the purpose here we will take a CFT to be determined by 
\begin{itemize}
\item all operators $\mathcal{O}_i$ with their {\em spin} $s$ ($s= 0,\tfrac{1}{2}, 1\ldots$) and  {\em scaling dimension} $\Delta_{i}$, and
\item the OPE coefficients $c_{ijk}$.
\end{itemize}
This is jointly called the CFT data:  $\{ (s_i,\Delta_i) , c_{ijk} \}$. This data defines a CFT assuming that it satisfies the mathematical consistency constraints of a conformal field theory. We now proceed to discuss one such key constraint used in the conformal bootstrap.

The 4-point correlators are not completely fixed by conformal symmetry, but we still have a certain handle of them. Suppressing the spacetime dependence, 
$\big \< \mathcal{O}_1 \mathcal{O}_2 \mathcal{O}_3 \mathcal{O}_4\big\>$ can be expressed via the OPE. For example, we can use the OPE on $\mathcal{O}_1 \mathcal{O}_2$ and $\mathcal{O}_3 \mathcal{O}_4$ or alternatively we could do with $\mathcal{O}_1 \mathcal{O}_4$ and $\mathcal{O}_2 \mathcal{O}_3$. The two expansions have to give the same result. Diagrammatically, we can illustrate this as 
\be
  \label{crossingsimple}
    \big \< \mathcal{O}_1 \mathcal{O}_2 \mathcal{O}_3 \mathcal{O}_4\big\>
    ~~=~~
    \sum_\mathcal{O} 
   \raisebox{-0.6cm}{
    \begin{tikzpicture}[scale=0.4, line width=1 pt]
    	\draw (-1,1)--(0,0);
	\draw (-1,-1)--(0,0);
	\draw (0,0)--(2,0);
	\draw (2,0)--(3,1);
	\draw (2,0)--(3,-1);
	\node at (-1.4,1) {$1$};
	\node at (-1.4,-1) {$2$};
	\node at (3.4,-1) {$3$};
	\node at (3.4,1) {$4$};
	\node at (1.2,0.55) {$\mathcal{O}$};
    \end{tikzpicture}
    }
   ~~=~~
   \sum_{\mathcal{O}'} 
   \hspace{-0.3cm}
      \raisebox{-1.cm}{
    \begin{tikzpicture}[scale=0.4, line width=1 pt]
    	\draw (-1,2)--(0,1);
	\draw (1,2)--(0,1);
	\draw (0,-1)--(0,1);
	\draw (-1,-2)--(0,-1);
	\draw (1,-2)--(0,-1);
	\node at (-1.4,2) {$1$};
	\node at (-1.4,-2) {$2$};
	\node at (1.45,-2) {$3$};
	\node at (1.45,2) {$4$};
	\node at (0.8,0.1) {$\mathcal{O}'$};
    \end{tikzpicture}
    }
  \,.      
\ee
This is called the {\em crossing relation}. 

The idea of the conformal bootstrap is to assume some CFT data and then subject them to the constraints of the crossing relation. It sounds perhaps too simple that the equivalence of two infinite sums can lead to significant and powerful constraints on the CFT data, but this is nonetheless the case. 

\subsection{Bootstrap of the 3d Ising Model}
Consider, as an example, a scalar field $\sigma(x)$ in a 3d CFT. It has spin 0 and  its  scaling dimension is classically equal to its mass-dimension: $\Delta_\sigma= \tfrac{1}{2}$. In a free theory, i.e.~with no interactions, the operator $\varepsilon(x) = (\sigma(x))^2$ has scaling dimension twice that of $\sigma(x)$ so $\Delta_\varepsilon= 1$. Now suppose quantum corrections in some putative interactive CFT makes $\Delta_\sigma= 0.53$. Now $\Delta_\varepsilon$ no longer has to be $2\Delta_\sigma$, it has a quantum life of its own. But what possible values can it take? Would it be realistic to think that a small quantum correction that increases $\Delta_\sigma$ from $0.5$ to $0.53$ allows $\Delta_\varepsilon$ to become as big as, say, 7? Our intuition says that this is unreasonable: if the correction to $\Delta_\sigma$ is small, then the correction to $\Delta_\varepsilon$ should also be small. Indeed, by analyzing the crossing relation \reef{crossingsimple} for the 4-point correlator $\<\sigma \sigma \sigma \sigma \>$, it can be shown numerically \cite{ElShowk:2012ht} that there exists no 3d CFTs with $\Delta_\sigma = 0.53$ and $\Delta_\varepsilon \gtrsim 1.45$.

The numerical implementation of the conformal bootstrap takes advantage of the fact that the crossing relation for $\<\sigma \sigma \sigma \sigma \>$ can be reformulated into a statement that schematically looks like
\be
  \label{schm}
  \sum_{\mathcal{O}} c_{\sigma\sigma\mathcal{O}}^2\, 
  \vec{v}_{\sigma\sigma\mathcal{O}} 
  ~=~ 0\,,
\ee
where $\vec{v}_{\sigma\sigma\mathcal{O}}$ represent vectors in a multi-dimensional abstract space (for more details, see Section \ref{s:confboot}). In a unitary CFT, the OPE coefficients $c_{\sigma\sigma\mathcal{O}}$ are real-valued so that means the coefficients in \reef{schm} are non-negative. Thus, consistency with the crossing relation requires that a sum of vectors with non-negative coefficients vanish. This is a non-trivial constraint. The key point then is that the functional form of the vectors $\vec{v}_{\sigma\sigma\mathcal{O}}$ is known and the only unknowns in \reef{schm} are the scaling dimensions $\Delta_\mathcal{O}$ that the $\vec{v}_{\sigma\sigma\mathcal{O}}$ depend on and the OPE coefficients $c_{\sigma\sigma\mathcal{O}}$. So for given input, say $\Delta_\sigma = 0.55$ in 3d, one can scan over values of $\Delta_\varepsilon$ and test numerically if there exists solutions to \reef{schm}. A ``no" means NO: there is no 3d CFT with scalar operators of those scaling dimensions. This is how the bound $\Delta_\varepsilon \gtrsim 1.45$ for  $\Delta_\sigma = 0.53$ was found in \cite{ElShowk:2012ht}.  A ``yes" means MAYBE: the putative CFT is not ruled out, but it does not mean it exists. One has to study higher dimensional operators and a broader set of 4-point correlators to find out if they are consistent with crossing too. So this analysis does not state whether or not there exists any 3d CFT with 
$\Delta_\sigma = 0.53$ and $\Delta_\varepsilon < 1.45$.\footnote{I'm glossing over much detail here. For example, the sum in \reef{schm} is over infinitely many operators, but it is controlled by knowing that the contributions from operators with large scaling dimension $\Delta_{\mathcal{O}}$ tend to be exponentially suppressed. There are also typically other assumptions made on the spectrum, for example such as the existence of a symmetry $\sigma \to -\sigma$ that implies that $\<\sigma\sigma\sigma\> = 0$ and hence $\sigma$ cannot appear in the $\sigma\sigma$ OPE. Further it is assumed that  $\varepsilon$ is the lowest-dimensional operator that can appear in the $\sigma\sigma$ OPE.} 

\begin{figure}[t!]
  \centerline{\includegraphics[width=11cm]{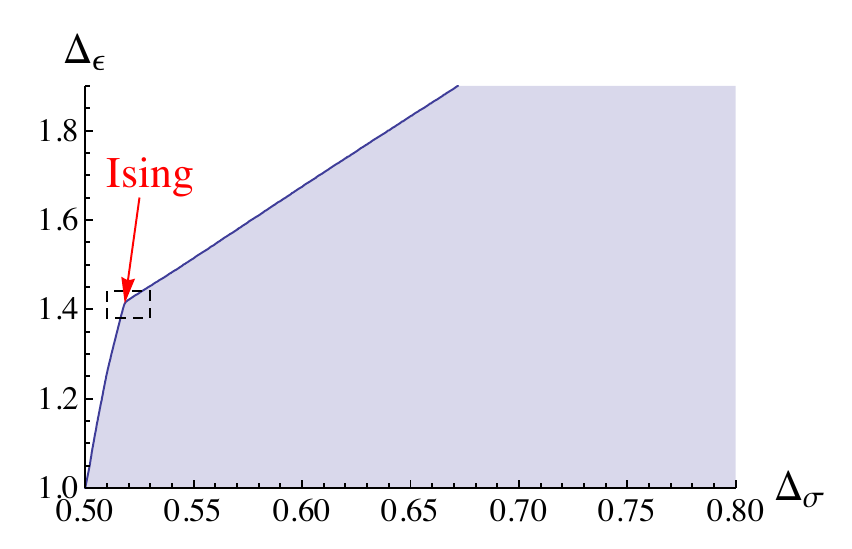}}
  \caption[Ising kink]{Plot showing the bounds on the scaling dimensions $\Delta_\sigma$ and  $\Delta_\varepsilon$ of the two lowest-dimension operators in a 3d CFTs with $\mathbb{Z}_2$ symmetry. The white region is excluded. This is based on numerical examination of the crossing constraints on a single correlator $\< \sigma\sigma\sigma\sigma\>$. The kink in the boundary between the excluded and non-excluded regions occurs near the expected location of the 3d Ising mode. This plot was originally presented  
  in \cite{ElShowk:2012ht} 
  by  El-Showk, Paulos, Poland,  Rychkov,  Simmons-Duffin, and  Vichi, and reproduced here with the permission of the authors.}
  \label{fig:kink} 
\end{figure}

When the crossing constraints are simultaneuosly applied to multiple correlators, the numerical bootstrap becomes even more powerful. For example, when applied \cite{Kos:2014bka} simultaneously to the three correlators   
$\<\sigma \sigma \sigma \sigma \>$, $\<\sigma \sigma \varepsilon \varepsilon \>$, and
$\<\varepsilon \varepsilon \varepsilon \varepsilon \>$, the conformal bootstrap with $\sigma \to - \sigma$ symmetry {\em rules out} any interacting 3d CFT with $\Delta_\sigma = 0.53$! While this is nice, there is an even more impressive and important result coming out of this analysis.

To back up, consider again the numerical bootstrap applied to a single $\<\sigma \sigma \sigma \sigma \>$. As a function of $\Delta_\sigma$, a scan over possible values of $\Delta_\varepsilon$ gives a bound $\Delta_\varepsilon < f(\Delta_\sigma)$ for some function $f$. The plot is shown in Figure  \ref{fig:kink}. The white region is rule out, the shaded region is not ruled out in this analysis. The key property to note here is the indication of a kink in the boundary curve near $\Delta_\sigma \approx 0.52$ and $\Delta_\varepsilon \approx 1.42$ \cite{ElShowk:2012ht}. Those are in fact close to the values expected for the two   lowest-dimension operators in the 3d Ising model!

\begin{figure}[t!]
  \centerline{\includegraphics[width=11cm]{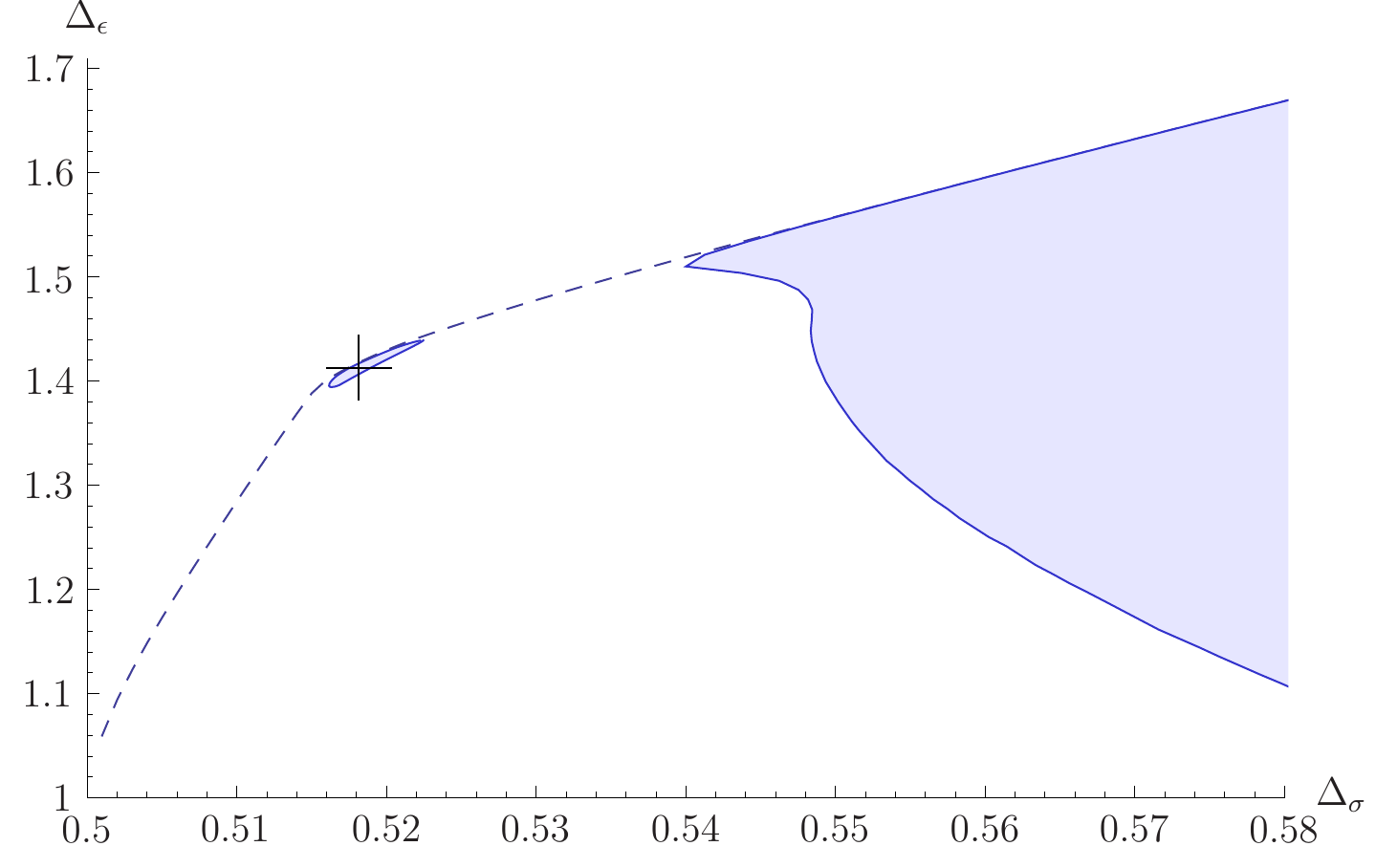}}
  \caption{Plot showing the excluded regions of scaling dimensions $\Delta_\sigma$ and  $\Delta_\varepsilon$ of the two lowest-dimension operators in a 3d CFTs with $\mathbb{Z}_2$ symmetry. This is based on numerical examination of the crossing constraints on three correlators: $\< \sigma\sigma\sigma\sigma\>$, $\< \sigma\sigma\varepsilon\varepsilon\>$, and $\< \varepsilon\varepsilon\varepsilon\varepsilon\>$. Where the kink occurred in Figure \ref{fig:kink}, there is now a small island of non-excluded theory-space, narrowing in on the 3d Ising mode. This plot was originally  presented 
  in \cite{Kos:2014bka} 
  by Kos,  Poland, and Simmons-Duffin, reproduced here with the authors' permission.}
  \label{fig:island}
\end{figure}

Now beefing up the analysis to apply the numerical bootstrap to the three correlators simultanously, it was found \cite{Kos:2014bka}  that a small island around the expected `location' of the 3d Ising model is cut out: see Figure \ref{fig:island}. This means that a large region of parameter space (white in the plot) is ruled out by the crossing constraints and there is a small shaded island-region not ruled out.

\begin{figure}[t!]
  \centerline{\includegraphics[width=11cm]{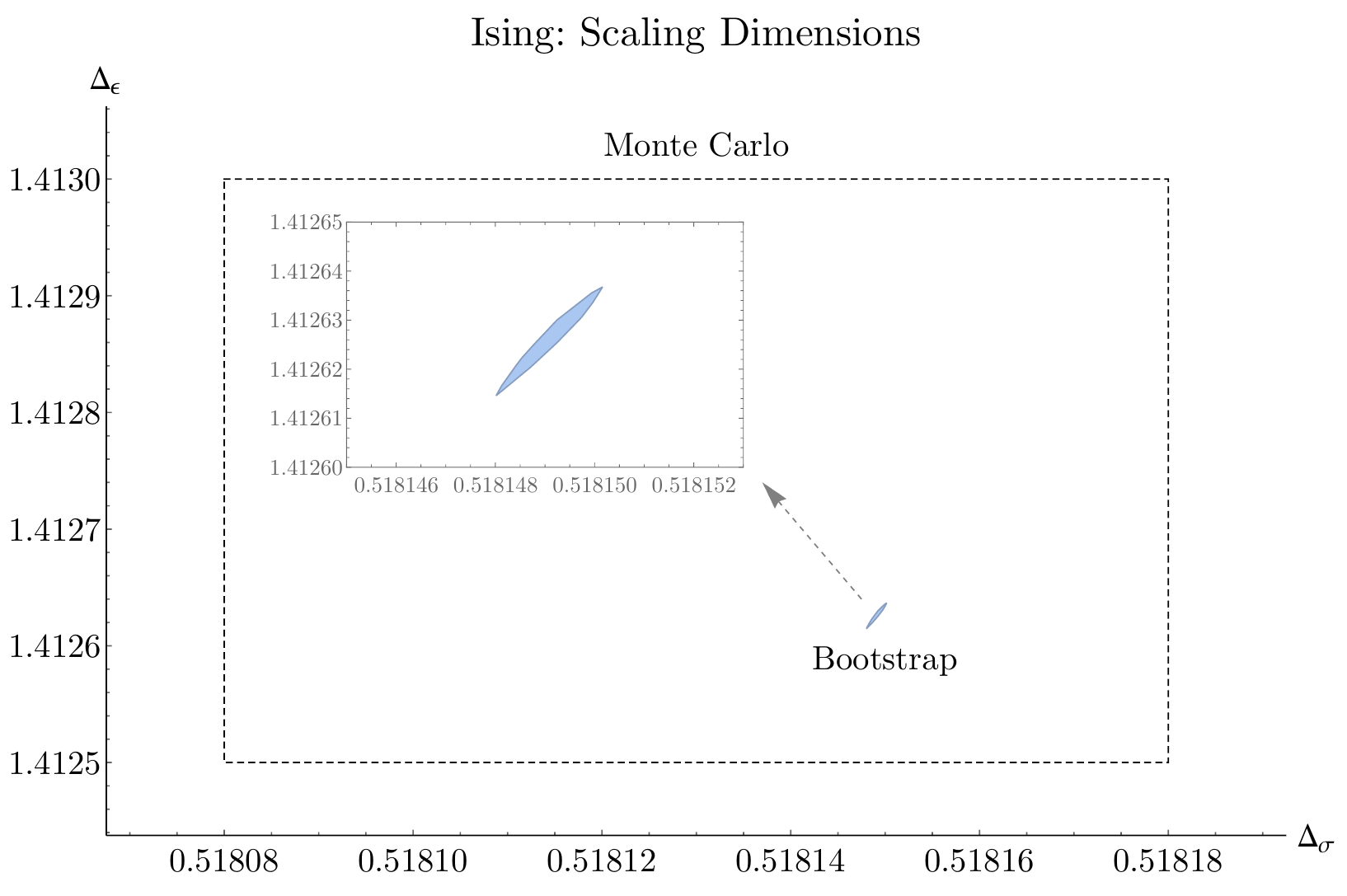}}
  \caption{Comparing the multi-correlator bootstrap results for the 3d Ising model versus Monte Carlo.  This plot was originally  presented 
  in \cite{Kos:2016ysd} by Kos,  Poland,  Simmons-Duffin, and Vichi, reproduced here with the permission of the authors. }
  \label{fig:bootMC}
\end{figure}

Precision numerics has made it possible to zoom in on this island and constrain it much further. The authors of \cite{Kos:2014bka,Kos:2016ysd}  used this to determine 
\be 
(\Delta_\sigma, \Delta_\varepsilon) = (0.5181489(10),1.412625(10)) \,,
\ee
which is higher precision that the available Monte Carlo results; see the comparison in Figure \ref{fig:bootMC}. Moreover, the numerical bootstrap determines the scaling dimensions and OPE coefficients to high precision of several of the lowest dimension operators in the 3d Ising model, not just of $\sigma$ and $\varepsilon$, see for example Table II in \cite{Poland:2018epd}. This offers evidence that the 3d Ising model may be a CFT, as proposed by Polyakov \cite{Polyakov:1970xd}.

Earlier in this presentation we mentioned the 3d Ising model: recall from Section \ref{s:watermagnet} that the critical point of water and ferromagnets is expected to be described by the 3d Ising model. The critical exponents of the approach to the critical point are directly related to the scaling dimensions $\Delta_\sigma$ and  $\Delta_\varepsilon$. For example, the critical exponent $\nu$ of the correlation length in \reef{xi} is determined by $\Delta_\varepsilon$ as
\be
   \nu = \frac{1}{d-\Delta_\varepsilon} \,.
\ee
Hence, for $d=3$ and the numerical bootstrap value of $\Delta_\varepsilon$, one finds $\nu = 0.62999(5)$ \cite{El-Showk:2014dwa}. Compare that with the experimental value of $\nu = 0.63$. It is clear that the formal theory exploration of the landscape of CFTs is relevant also for observable physics in our world.

The idea of the conformal bootstrap dates back about 50 years \cite{Ferrara:1973yt,Polyakov:1974gs}, but it had a revival starting around 2008 when problem was phrased in terms of bounding operators and especially when it was realized that the condition \reef{crossingsimple} can be phrased a form that makes it particularly well-suited for numerical implementation. 
Initial work was driven by particle physics applications (see e.g.~\cite{Rattazzi:2008pe} and \cite{Poland:2010wg})  and the recent application to the 3d Ising model was initiated in \cite{Rychkov:2011et,ElShowk:2012ht}. 
Practically, this means that the crossing relation is rephrased a statement about whether a set of vectors can add to zero using only non-negative coefficients. Semidefinite programming techniques \cite{Simmons-Duffin:2015qma} have turned out to be powerful approaches to asses such questions numerically. By now, quite a number of analytical results have also be developed to further enhance the exploration of the CFT landscape.  

\vspace{3mm}
\noindent {\bf Other examples: Helium, QED$_3$, SCFTs Across Dimensions}\\[1mm]
There have been a multitude of applications of --- and other developments closely related to --- the conformal bootstrap. 

A very nice example are 3d CFTs with $O(N)$ global symmetry studied using the conformal bootstrap in \cite{Kos:2015mba,Kos:2016ysd}. In particular, the $O(2)$ model that is expected to describe the $\lambda$-line superfluid transition in Helium-4. The physics of this critical transition lies in a different universality class (sometimes also called the 3d XY universality class) from that of the liquid-vapor critical point that we described in Section \ref{s:watermagnet} for water and the Curie point of ferromagnets. The difference is manifest in the values of the critical exponents such as $\nu$ associated with the divergence of the correlation length: for the liquid-vapor critical point $\nu = 0.63\ldots$, but for the $\lambda$-line transition it is approximately $\nu = 0.67\ldots$. 
Experimental results \cite{Lipa:2003zz}, Monte Carlo simulations \cite{Campostrini_2006,Hasenbusch_2019}, and the conformal bootstrap methods \cite{Chester:2019ifh} agree on the value of $\nu$ to 2-decimal places. The theoretical methods  are in agreement beyond the leading digits within the uncertainties, with the conformal bootstrap giving the highest precision result \cite{Chester:2019ifh,Ryckkovpage}
\be
\nu = 0.67175(10) \,.
\ee
However, there is an 8$\sigma$ tension between the theoretical values of $\nu$ and the experimental one \cite{Campostrini_2006,Chester:2019ifh,Ryckkovpage}.

There are many other applications of the conformal bootstrap. Supersymmetric CFTs have been proposed to be relevant in the context of topological superconductors and for the description of a critical point on the surface of topological insulators.  Another class of 3d CFTs arise as fixed points of 3d models with gauge fields. Two main classes have been examined: 3d Chern-Simons fields coupled conformally to matter and bosonic QED$_3$ models. A nice overview of these topics are given in the review \cite{Poland:2018epd}. 

There are several explorations of 4d CFTs as well as in 5d and 6d. The theories of particular interest to explore with the conformal bootstrap are those without Lagrangian descriptions, for example the 6d  (2,0) supersymmetric CFT of M5 branes \cite{Beem:2015aoa} or the non-Lagrangian 
$\mathcal{N}=2$ supersymmetric CFTs in 4d \cite{Beem:2014zpa}. We mentioned in footnote \ref{footieN3} that Lagrangian $\mathcal{N}=3$ SYM was equivalent to the $\mathcal{N}=4$ SYM theory. However, non-Lagrangian theories with $\mathcal{N}=3$ theories without $\mathcal{N}=4$ SUSY are not ruled out \cite{Aharony:2015oyb,Garcia-Etxebarria:2015wns} and the conformal bootstrap techniques have been applied to place bounds on the space of $\mathcal{N}=3$ SCFTs \cite{Lemos:2016xke}. 
The conformal bootstrap has also been applied in conjunction with another modern technique, supersymmetric localization, to derive results about the M2-brane theory in M-theory \cite{Agmon:2017xes}. 

The conformal bootstrap is a powerful method with wide applicability. In one form, its numerical implementation has resulted in new results not only to explore the space of possible CFTs, but also to pinpoint the properties of known ones, such as the critical exponents. There are very nice reviews about CFTs and the conformal bootstrap, see for example \cite{Poland:2016chs},  \cite{Simmons-Duffin:2016gjk}, \cite{Rychkov:2016iqz}, \cite{Poland:2018epd}, and \cite{Chester:2019wfx}. The presentations here and in Section \ref{s:confboot} rely heavily on those references. 

\section{Technical: Amplitude Bootstrap}
\label{s:ampbootex}

In this section, we revisit the question posed in the beginning of Section \ref{s:ampboot} and address it in full technical detail:

\begin{quote}
 {\em ``Does there exist a local relativistic QFT with two massless real scalars such that every tree amplitude vanishes in the limit where a single momentum is taken soft?  Is such a model unique? Must it have any particular symmetries, such as an interchange symmetry that requires the scalar particles to be on equal footing?".} 
\end{quote}

Let us begin by phrasing the question in the traditional QFT languae, i.e.~in terms of fields and symmetries of a Lagrangian. First, it is convenient to combine the two real scalar fields $\phi_1$ and $\phi_2$ into a complex scalar 
\be
   Z = \phi_1 + i \phi_2\,,~~~~~~
   \bar{Z} = \phi_1 - i \phi_2\,.
\ee 
A canonical kinetic term can  be written $\partial_\mu Z \partial^\mu \bar{Z}$. In the bottom-up Lagrangian approach, one assumes such a kinetic term and then builds up interaction terms. 

Secondly, the vanishing soft limit is related to (but not identical to) the Lagrangian having a shift symmetry $Z \to Z + c + \ldots$, where $c$ is a complex-valued constant and the ellipses stand for field-dependent terms. (For a more precise statement of this relation, see Section \ref{s:coset}.) Models in which there is at least one derivative on every scalar field trivially have such a shift symmetry and also have amplitudes with  vanishing soft limits. For example interactions like $(\partial Z)^4$ gives Feynman rules in which each term is a products of the four momenta and therefore the vertex vanishes when any one of them is taken soft. This trivial way of realizing the shift symmetry gives contributions at  $O(p^4)$ to amplitudes, so the question here really is if there are models that realize the symmetries at lower order in the low-energy expansion. With no derivatives on the fields in the interactions, the model cannot have a shift symmetry or vanishing soft limits. A 2-derivative theory gives $O(p^2)$ amplitudes and it could have 4-point interaction such as $Z^2 (\partial {Z})^2$ and similar with $\bar{Z}$'s too.
 Such terms do not obviously have a shift symmetry or give amplitudes with vanishing soft limits. So we have to add other interaction terms to achieve this and the question is how much freedom there is to do so. 
 
Thirdly, the question of whether the two real scalars $\phi_1$ and $\phi_2$ are on equal-footing translates to whether the model has an $SO(2)$ symmetry that acts on $(\phi_1,\phi_2)$ as a rotation. In the language of the complex scalar, this is equivalent to asking if there is a $U(1)$ symmetry that 
takes  $Z \to e^{i\alpha}Z$ and  $\bar{Z} \to e^{-i\alpha} \bar{Z}$ for some constant $\alpha$. We do {\em not} assume a $U(1)$ symmetry, but will see that it emerges in the leading-order model. 

Rather than attempting to build a Lagrangian whose Feynman rules give amplitudes that vanish in the soft limit, the modern on-shell amplitude approach starts with the amplitudes to systematically determined what models can be ruled out and map out which ones may exist. 

\subsection{Setup}

We rephrase the problem in terms of on-shell amplitudes.  

\vspace{2mm}
\noindent {\bf Variables}.
 We consider scattering of $n$ massless particles so the external 4-momenta $p_i^\mu$, where $i=1,2,3,\dots, n$, are required to satisfy the on-shell condition $p_i^2  = p_{i\mu} p_i^\mu = 0$.  
 Momentum conservation requires\footnote{The conservation of momentum is really that the sum of incoming momenta must equal the sum of outgoing momenta. As a technical tool, crossing symmetry is often used to trade in-coming particles for outgoing ones, so that the amplitude has all external particles on equal footing as outgoing. This is helps make the calculations a bit simpler and once a result is obtained, particles can simply the ``crossed" back to let some of them be incoming again for the purpose of computing the cross-section.}
 \be
  \label{mymomcons}
   \sum_{i=1}^n p_i^\mu = 0\,.  
\ee
The amplitude must depend on the external momenta in a Lorentz invariant way, namely as dot-products $p_i.p_j = p_{i\mu} p_j^\mu$, where $i$ is the particle label $i=1,2,3,\dots, n$.   Since $p_i^2=0$ for all $i$, we have
\be
   s_{ij} \equiv (p_i + p_j)^2 =  2 p_i.p_j \,,~~~~~~~
   s_{ijk} \equiv (p_i + p_j+p_k)^2 =  2 p_i.p_j +2 p_i.p_k+2 p_j.p_k 
   \,,~~~~\text{etc.}
\ee
The Mandelstam variables $s_{ij\ldots}$ are not all independent due to the constraints of momentum conservation \reef{mymomcons}. For example, for $n=4$, we have
\be
  \label{n4momcons}
 n=4\text{:}~~~~~~
 s_{12}=s_{34}\,,~~~~
 s_{13}=s_{24} \,,~~~~
 s_{14}=s_{23}\,,~~~~
 \text{and}~~~~
 s_{12} + s_{13}+ s_{14} = 0\,.
\ee 
Thus the scalar amplitudes are functions of Mandelstam variables  subject to the constraints of momentum conservation. 

\vspace{2mm}
\noindent {\bf Analytic Structure.} Tree amplitudes must be  rational functions of Mandelstam variables. In a local theory of massless scalars, they can have simple poles (and no higher-order poles) at locations where $s_{ij\ldots k} \to 0$ and the residue of such a pole is, by unitarity, a product of lower-point amplitudes. There can also be polynomial terms in the amplitude; in an $n$-point amplitude such polynomial terms can arise from $n$-point interactions in the Lagrangian. 

\vspace{2mm}
\noindent {\bf Bose Symmetry.} Bose symmetry requires that the amplitude is invariant under exchanges of identical bosons.
Specifically, $A_n(Z \dots Z \bar{Z} \dots \bar{Z})$ must be symmetric under all permutations of the momenta of the $Z$'s, and likewise for those of the $\bar{Z}$'s

\vspace{2mm}
\noindent {\bf Vanishing Single Soft Limits}. 
As any one of the external momenta is taken to zero, the amplitude has to vanish in the momentum:
\be
 \label{softlimit}
  A_n(Z \dots Z \bar{Z} \dots \bar{Z}) \to 0 ~~~\text{for any}~~~p_i\to 0\,.
\ee
As noted in Section \ref{s:ampboot}, this is called a vanishing soft theorem or an Adler zero \cite{Adler:1964um}.

\vspace{2mm}
\noindent {\bf $U(1)$ Symmetry}. The $U(1)$ symmetry can be understood as the statement that $Z$ particles have charge $+1$ and $\bar{Z}$ particles have charge $-1$. The associated Ward identity then says 
\be 
    A_n(\underbrace{Z \dots Z}_{n_Z} \underbrace{\bar{Z} \dots \bar{Z}}_{n_{\bar{Z}}}) ~=~ 0   
    ~~~~\text{for}~~~~
    n_Z \ne n_{\bar{Z}}\,.
\ee
We do not assume $U(1)$ symmetry, we shall see it emerge in the leading order theory in the sense that it only has non-vanishing amplitudes with $n_Z = n_{\bar{Z}}$.

\vspace{2mm}
The goal is to construct the model subject to the constraints of vanishing soft limit at the lowest possibles order in the energy-momentum expansion; higher order terms are considered in Section \ref{s:hdcorrections}. To streamline the discussion, we first look at amplitudes that violate the $U(1)$, then those that respect it. 

\subsection{Bootstrapping the Model:  $U(1)$-Violating Amplitudes}
\label{s:cp1bootnoU1}

Amplitudes that do not obey the $U(1)$ symmetry are those with an unequal number of $Z$ and $\bar{Z}$'s. 
\begin{itemize}

\item 
{\bf 3-point.} Momentum conservation for 3 massless particles sets all Mandelstams to zero, e.g.~$s_{12} = (p_1+p_2)^2 = p_3^2 = 0$. So there can be no momentum-dependence in the 3-particle scalar amplitudes, they have to be constants. For example, $A_3(ZZZ) = d_0$ and $A_3(ZZ\bar{Z}) = d_0'$. But this is at odds with the assumption of vanishing soft behavior \reef{softlimit}.\footnote{The special kinematics associated with taking soft limits of a 3-particle amplitude may appear tricky; however, a constant 3-particle amplitude necessarily implies a divergence of the 4-particle amplitudes via pole diagrams \cite{Elvang:2016qvq}, so the requirement of a vanishing soft limit rules out the 3-particle amplitudes.}
We conclude that there can be no 3-point amplitudes. 

\item 
{\bf 4-point.} At 4-points, the $U(1)$ violating amplitudes are
$A_4(Z Z Z Z )$, $A_4(Z Z Z \bar{Z} )$, and their conjugates. They cannot have poles, since there are no 3-particle amplitudes they could factor into (unitarity), so they have to be polynomial in the Mandelstam variables. Constant terms are excluded by the soft limit \reef{softlimit}. At $O(p^2)$, the only Mandelstam polynomial compatible with the Bose symmetry of 3 or 4 identical scalars is $s_{12}+s_{13}+s_{23}$, but that vanishes by momentum conservation \reef{n4momcons}. Hence $A_4(Z Z Z Z )$  or $A_4(Z Z Z \bar{Z} )$ start at $O(p^4)$. 

\item 
{\bf 5-point.} In the absence of 3-particle amplitudes, the 5-particle amplitudes cannot have poles, so they must be polynomial in the Mandelstam variables. A constant is incompatible with the vanishing soft limit.  At order $O(p^2)$, Bose symmetry requires 
$A_5(ZZ Z ZZ)$ and $A_5(ZZ Z Z\bar{Z} )$ to be $\sum_{1<i<j<5} s_{ij}$, but this is zero due to momentum conservation. The amplitude with three $Z$'s and two $\bar{Z}$'s is uniquely determined at $O(p^2)$ to be $A_5(ZZ Z \bar{Z}\bar{Z} ) = a  s_{45}$, but this does not vanish in the limit of taking (say) $p_1 \to 0$. So we must set $a=0$ and hence 5-point amplitudes are at least  $O(p^4)$.

\item {\bf 6-point and above.} Just as in the 5-point case, any amplitude with all $Z$'s or a single $\bar{Z}$ vanishes at $O(p^2)$. With at least two of both  $Z$ and $\bar{Z}$, there is a unique  Bose symmetric Mandelstam polynomial at $O(p^2)$, namely $s_{12\ldots n_Z}$ where we have chosen the $Z$ particles to have momentum labels $i=1,2,\ldots,n_Z$. However, this is non-vanishing in the soft limits of any $\bar{Z}$ momenta.
\end{itemize}

Based on the above, we conclude that  any $U(1)$-violating amplitude obeying the vanishing soft limit condition   have to be at least $O(p^4)$. Such higher-order corrections are considered in Section \ref{s:hdcorrections}. The lesson here is that any complex scalar 2-derivative (i.e.~$O(p^2)$ amplitudes)  model with vanishing single soft limits {\em must} have $U(1)$ symmetry: this is an example of a --- perhaps surprising --- emergent symmetry.

\subsection{Bootstrapping the Model: $U(1)$-Conserving Amplitudes}
\label{s:cp1boot}

Amplitudes with $U(1)$ symmetry have an equal number of $Z$ and $\bar{Z}$'s, so in particular they must be even-point. 
\begin{itemize}
\item 
{\bf 4-point $U(1)$ conserving.} 
As the lowest-point amplitude $A_4(Z \bar{Z} Z \bar{Z})$ must be polynomial in the Mandelstam variables. A constant is excluded by the vanishing soft limit constraint, so we write as most general ansatz with the appropriate Bose symmetry:
\be
  A_4(Z \bar{Z} Z \bar{Z} ) = \frac{a_1}{\Lambda^2} \, s_{13} 
  +O(p^4)\,.
\ee
At order $O(p^2)$, the polynomial $s_{12}+s_{23} = s_{34}+s_{14}$ is also compatible with Bose symmetry, but by momentum conservation \reef{n4momcons} is equal to $-s_{13}$. We have included a scale $\Lambda$ of mass-dimension $1$ so that $a_1$ is a pure number.   The amplitude vanishes in the soft limit of any one of the momenta, so there are no constraints on $a_1$.

\item 
{\bf 6-point.} The argument in  Section \ref{s:cp1bootnoU1} shows that any polynomial term at $O(p^2)$ in a scalar amplitude with more than 4 external legs is non-vanishing in the single soft limit. This means that to achieve a vanishing soft limit beyond 4-points, there must be cancellations among the pole-terms and the contact terms. 

\vspace{0.8mm}
At 6-points the pole terms must have residues that are products of two 4-point amplitudes, for example on the 123-channel:
\be 
  \label{poleterm}
     \raisebox{-0.58cm}{
    \begin{tikzpicture}[scale=0.4, line width=1 pt]
    	\draw (-1,1)--(0,0);
	\draw (-1,-1)--(0,0);
	\draw (-1.5,0)--(3.5,0);
	\draw (2,0)--(3,1);
	\draw (2,0)--(3,-1);
	\node at (-1.4,1) {\small $1$};
	\node at (-1.9,0) {\small $\bar{2}$};
	\node at (-1.4,-1) {\small$3$};
	\node at (3.5,-1) {\small$\bar{4}$};
	\node at (3.9,0) {\small $5$};
	\node at (3.5,1) {\small $\bar{6}$};
	\node at (1.17,0.55) {\small $P$};
    \end{tikzpicture}
    }
  =~ \frac{A_4(Z_1 \bar{Z}_2 Z_3 \bar{Z}_P ) A_4(Z_{-P} \bar{Z}_4 Z_5 \bar{Z}_6 )}{s_{123}} = \frac{a_1^2}{\Lambda^4} \frac{s_{13} s_{46}}{s_{123}} \,.
\ee
The most general contact terms can be parameterized as
\be
  \label{CT6}
       \raisebox{-0.58cm}{
    \begin{tikzpicture}[scale=0.4, line width=1 pt]
    	\draw (-1,1)--(0,0);
	\draw (-1,-1)--(0,0);
	\draw (-1.5,0)--(1.5,0);
    	\draw (1,1)--(0,0);
    	\draw (1,-1)--(0,0);
	\node at (-1.4,1) {\small $1$};
	\node at (-1.9,0) {\small$\bar{2}$};
	\node at (-1.4,-1) {\small$3$};
	\node at (1.5,-1) {\small$\bar{4}$};
	\node at (1.9,0) {\small $5$};
	\node at (1.5,1) {\small $\bar{6}$};
    \end{tikzpicture}
    }
  = ~b_0 + b_1 \,s_{246} + O(p^4)\,.
\ee
There are no other independent polynomials with Bose symmetry at  $O(p^2)$.
So we can write the 6-point amplitude as the sum of all the pole diagrams and the possible contact terms:
\be
  \label{A6}
  A_6(Z \bar{Z}  Z \bar{Z}Z \bar{Z} ) = 
  \frac{a_1^2}{\Lambda^4} \bigg( 
  \frac{s_{13} s_{46}}{s_{123}} 
  +\frac{s_{13} s_{26}}{s_{143}}  
  +\frac{s_{13} s_{24}}{s_{163}} + (1 \lra 5)+ (3 \lra 5)  \bigg)
  + b_0 + b_1\, s_{246} + O(p^4)\,.
\ee
In the soft limit $p_6 \to 0$, the first two pole terms in  \reef{A6} vanish while the third one reduces to $s_{24}$. Under the exchanges $(1 \lra 5)$ and $(3 \lra 5)$ two more $s_{24}$'s are generated. So we get 
\be
  A_6(Z \bar{Z}  Z \bar{Z}Z \bar{Z} )  \to 
  3 \frac{a_1^2}{\Lambda^4} s_{24}
  + b_0 + b_1 s_{24} + O(p^4)\,.
\ee
Therefore, in order to have vanishing soft limits \reef{softlimit} at 6-points, we must have 
\be
  \label{b1fixed}
  b_0=0 
  ~~~~\text{and}~~~~
  b_1 =- 3 \frac{a_1^2}{\Lambda^4}\,.
\ee 
Hence,  the 4-point and 6-point amplitudes are completely fixed at order $O(p^2)$ in terms of just one number, the coupling constant $a_1$.

\item 
{\bf  8-point.}
The above pattern continues to higher-point amplitudes: at $O(p^2)$ the whole model is uniquely fixed by the symmetry requirements in terms of a number, $a_1$, and a single dimensionful scale $\Lambda$. 

\vspace{0.8mm}
Consider at 8-points, the $p_8 \to 0$ soft limit. Some diagrams directly vanish in this limit. Of diagrams that do not vanish, those that result in terms with poles directly vanish among themselves. To see this, consider the three diagrams with a $1/s_{123}$ pole that do not vanish in the $p_8 \to 0$ limit: 
\bea
   \label{8ptA}
      \hspace{-1cm} \phantom{33} \raisebox{-0.78cm}{
    \begin{tikzpicture}[scale=0.4, line width=1 pt]
    	\draw (-1,1)--(0,0);
	\draw (-1,-1)--(0,0);
	\draw (-1.5,0)--(3.5,0);
	\draw (2,0)--(3.2,0.9);
	\draw (2,0)--(3.2,-0.9);
	\draw (2,0)--(2.5,1.3);
	\draw (2,0)--(2.5,-1.3);
	\node at (-1.4,1) {\small $1$};
	\node at (-1.9,0) {\small$\bar{2}$};
	\node at (-1.4,-1) {\small$3$};
	\node at (2.9,1.5) {\small $\bar{8}$};
	\node at (3.6,-1) {\small$7$};
	\node at (3.9,0) {\small $\bar{6}$};
	\node at (3.6,1) {\small $5$};
	\node at (2.9,-1.5) {\small $\bar{4}$};
    \end{tikzpicture}
    }
    &=&~ \frac{\big(\frac{a_1}{\Lambda^2}s_{13}\big)\big(-\tfrac{3a_1^2}{\Lambda^4}s_{468}\big)}{s_{123}} 
   ~\to~ -3 \frac{a_1^3}{\Lambda^6} \frac{s_{13} s_{46}}{s_{123}}
   \,,
\\
   \label{8ptB}
\hspace{-1cm}
     \raisebox{-0.83cm}{
    \begin{tikzpicture}[scale=0.4, line width=1 pt]
    	\draw (-1,1)--(0,0);
	\draw (-1,-1)--(0,0);
	\draw (-1.5,0)--(5.5,0);
	\draw (4,0)--(5,1);
	\draw (4,0)--(5,-1);
	\draw (2,-1)--(2,1);
	\node at (-1.4,1) {\small $1$};
	\node at (-1.9,0) {\small $\bar{2}$};
	\node at (-1.4,-1) {\small$3$};
	\node at (5.5,-1) {\small${5}$};
	\node at (5.9,0) {\small $\bar{8}$};
	\node at (5.5,1) {\small ${7}$};
	\node at (2,1.6) {\small $\bar{4}$};
	\node at (2,-1.6) {\small $\bar{6}$};
    \end{tikzpicture}
    }
    &=&~ \frac{\big(\frac{a_1}{\Lambda^2}s_{13}\big)
    \big(\frac{a_1}{\Lambda^2}s_{46}\big)
    \big(\frac{a_1}{\Lambda^2}s_{57}\big)}{s_{123}\,s_{578} }
   ~\to~ \frac{a_1^3}{\Lambda^6} \frac{s_{13} s_{46}}{s_{123}}
   \,,
\eea
and
\be
   \label{8ptC}
   \begin{split}
     \raisebox{-0.83cm}{
    \begin{tikzpicture}[scale=0.4, line width=1 pt]
    	\draw (-1,1)--(0,0);
	\draw (-1,-1)--(0,0);
	\draw (-1.5,0)--(5.5,0);
	\draw (4,0)--(5,1);
	\draw (4,0)--(5,-1);
	\draw (2,-1)--(2,1);
	\node at (-1.4,1) {\small $1$};
	\node at (-1.9,0) {\small $\bar{2}$};
	\node at (-1.4,-1) {\small$3$};
	\node at (5.5,-1) {\small$\bar{4}$};
	\node at (5.9,0) {\small $5$};
	\node at (5.5,1) {\small $\bar{6}$};
	\node at (2,1.6) {\small $7$};
	\node at (2,-1.6) {\small $\bar{8}$};
    \end{tikzpicture}
    }
    ~+~(5 \lra 7)
    &~=~ \frac{\big(\frac{a_1}{\Lambda^2}s_{13}\big)
    \big(\frac{a_1}{\Lambda^2}s_{4568}\big)
    \big(\frac{a_1}{\Lambda^2}s_{46}\big)}{s_{123}\,s_{456} }
    ~+~(5 \lra 7)\\[-2mm]
   &~\to ~2 \frac{a_1^3}{\Lambda^6} \frac{s_{13} s_{46}}{s_{123}}
\,.
   \end{split}
\ee
The three contributions \reef{8ptA}, \reef{8ptB}, and \reef{8ptC} cancel, and it is clear why: this is guaranteed by the vanishing of the 6-point in the soft limit.

\vspace{0.8mm}
Finally there are pole diagrams with non-vanishing soft limits that do no cancel among themselves but leave behind polynomial terms: for example 
\be
        \raisebox{-0.78cm}{
    \begin{tikzpicture}[scale=0.4, line width=1 pt]
    	\draw (-1,1)--(0,0);
	\draw (-1,-1)--(0,0);
	\draw (-1.5,0)--(3.5,0);
	\draw (2,0)--(3.2,0.9);
	\draw (2,0)--(3.2,-0.9);
	\draw (2,0)--(2.5,1.3);
	\draw (2,0)--(2.5,-1.3);
	\node at (-1.4,1) {\small $7$};
	\node at (-1.9,0) {\small$\bar{8}$};
	\node at (-1.4,-1) {\small$1$};
	\node at (2.9,1.5) {\small $\bar{6}$};
	\node at (3.6,-1) {\small$3$};
	\node at (3.9,0) {\small $\bar{4}$};
	\node at (3.6,1) {\small $5$};
	\node at (2.9,-1.5) {\small $\bar{2}$};
    \end{tikzpicture}
    }
    = ~\frac{\big( \frac{a_1}{\Lambda^2}\big) s_{17}\big(-3\frac{a_1^2}{\Lambda^4}s_{246}\big)}{s_{178}} 
   ~\to~ -3 \frac{a_1^3}{\Lambda^6} s_{246} 
   ~~~~\text{as}~~~p_8 \to 0\,.
\ee
There are 6 distinct such diagrams (from the 6 possible pairings of odd-numbered momenta 1357 on the LHS), so that means that 
\be
  A_8(Z \bar{Z}  Z \bar{Z}Z \bar{Z} Z \bar{Z}) \to  -18 \frac{a_1^3}{\Lambda^6}   
  s_{246} ~~~~\text{as}~~~ p_8 \to 0\,.
\ee 
This can be canceled by an 8-point local contact term $+18\tfrac{a_1^3}{\Lambda^6}  s_{2468}$ in order to ensure the vanishing soft theorem.
\item
{\bf 10-point.} The pattern continues. The local terms that arise in the soft limit $p_{10} \to 0$ come from the pole diagrams with the 8-point contact terms, so it gives $18 \tfrac{a_1^4}{\Lambda^8} s_{2468}$. There are (5 choose 2)=10 such diagrams, so the $p_{10}$ soft limit of all the pole terms can be cancelled by a contribution $-180 \tfrac{a_1^4}{\Lambda^8} s_{2468\,10}$ from a local 10-point interaction. 

\end{itemize}
To summarize, we have found that the assumptions have fixed the amplitudes in the theory completely in the leading order of the low-energy expansion: the lowest-point non-vanishing amplitudes are at order $O(p^2)$ 
\be
  \begin{split}
  A_4(Z \bar{Z} Z \bar{Z} ) &=  \frac{1}{\Lambda^2} \, s_{13} 
  \,,
  \\
    A_6(Z \bar{Z}  Z \bar{Z}Z \bar{Z} ) &= 
  \frac{1}{\Lambda^4} \bigg( 
  \frac{s_{13} s_{46}}{s_{123}} 
  +\frac{s_{13} s_{26}}{s_{143}}  
  +\frac{s_{13} s_{24}}{s_{163}} + (1 \lra 5)+ (3 \lra 5)  
  -3s_{246}\bigg)
 \,,
 \\
    A_8(Z \bar{Z}  Z \bar{Z}Z \bar{Z} Z \bar{Z} ) &= \frac{1}{\Lambda^6}\bigg(\text{pole terms} +18  s_{2468}\bigg) 
    \,,
 \\
    A_{10}(Z \bar{Z}  Z \bar{Z}Z \bar{Z} Z \bar{Z}Z \bar{Z} ) &= \frac{1}{\Lambda^8}\bigg(\text{pole terms} -180  s_{2468\,10}\bigg) 
    \,.
  \end{split}
\ee
Here we have set the 4-point coupling $a_1=1$ without any loss of generality. 

For those who love Lagrangians, one can retro-engineer the interaction terms based on the polynomial terms above to find 
\be
  \label{CP1lag}
  \begin{split}
  \mathcal{L} = -\pa_\mu Z \pa^\mu \bar{Z} 
  &+  \frac{1}{\Lambda^2}Z \bar{Z}\pa_\mu Z \pa^\mu \bar{Z}
  -\frac{3}{4} \frac{1}{\Lambda^4}Z^2 \bar{Z}^2\pa_\mu Z \pa^\mu \bar{Z} 
  \\
  &
  + \frac{1}{2}\frac{1}{\Lambda^6}Z^3 \bar{Z}^3\pa_\mu Z \pa^\mu \bar{Z}
  - \frac{5}{16}\frac{1}{\Lambda^8}Z^4 \bar{Z}^4\pa_\mu Z \pa^\mu \bar{Z}
  + \ldots \,,
  \end{split}
\ee
where the dots stand for interactions with more than 10 fields. Here we have used that the $(2n)$-point matrix element of 
\be  
  \label{2nptint}
  Z^{n-1} \bar{Z}^{n-1} \pa_\mu Z \pa^\mu \bar{Z} 
  = \frac{1}{n^2} (\pa_\mu Z^n)(\pa^\mu \bar{Z}^n)
\ee
is
\be
  \label{2nptFeyn}
   \frac{1}{n^2} (n!)^2 \,
   i^2\bigg( \sum_{i~\text{odd}} p_i \bigg)
    \bigg( \sum_{j~\text{even}} p_j \bigg)   
   ~=~
   \big[(n-1)!\big]^2 s_{246 \dots 2n} \,,
\ee
using momentum conservation.
This was used to normalize the interaction terms so they exactly reproduce the local terms in the amplitudes above. For example, for the 10-point term in \reef{CP1lag},  the overall numerical factor of the local contact term contribution is $- \frac{5}{16} (4!)^2 = -180$. 

We can extent this to reasoning to $2n$-points. Suppose the numerical coefficient of the $2n$-point interaction term in the Lagrangian is $\a_n$; .e.g.~$\a_4= \tfrac{1}{2}$. Then from \reef{2nptint} and \reef{2nptFeyn}, the polynomial term it contributes to the amplitude is 
$\a_n \big[(n-1)!\big]^2 s_{246 \dots 2n}$. On the other hand, the purpose of this term will be to cancel the local contribution from the soft limit of the ($n$ choose 2) pole diagrams involving the $2(n-1)$-point local contribution, which has coefficient $\a_{n-1} \big[(n-2)!\big]^2$.
So we see that 
\be 
  \label{alpharecrel}
  \a_n \big[(n-1)!\big]^2 = -\binom{n}{2} \, \a_{n-1} \big[(n-2)!\big]^2
  ~~~~\implies~~~~
  \a_n 
     = -\frac{1}{2} \frac{n}{(n-1)}\a_{n-1} \,.
\ee
With $\a_1 = -1$ (the kinetic term), we get $\a_2 = 1$ and likewise we reproduce the numerical coefficients of other terms in \reef{CP1lag}. The recursive formula \reef{alpharecrel} is straightforward to solve and one finds
\be
   \a_n = (-1)^n \frac{n}{2^{n-1}} \,.
\ee
These are exactly the series coefficients of $1/{(1+\tfrac{1}{2}Z \bar{Z})^2}$ expanded around 
 zero!
Thus, including the scale $\Lambda$, we have discovered that the Lagrangian can be re-summed to 
\be
  \label{CP1Lag}
  \mathcal{L} = -\frac{\pa_\mu Z \pa^\mu \bar{Z}}{\big(1+\tfrac{1}{2\Lambda^2}Z \bar{Z}\big)^2}
   = -G_{Z\bar{Z}} \pa_\mu Z \pa^\mu \bar{Z}\,,
\ee 
where $G_{Z\bar{Z}} = 1/{(1+\tfrac{1}{2\Lambda^2}Z \bar{Z})^2}$ can be identified as the Fubini-Study K\"ahler metric on $\mathbb{CP}^1$. This 2-scalar model is well-known, it is the $\mathbb{CP}^1$ nonlinear sigma model (NLSM). It describes two real Goldstone modes arising from the spontaneous symmetry breaking of $SU(2)$ to $U(1)$. The real Goldstones are paired into the complex scalar $Z$ that ``lives" in the symmetric coset space $SU(2)/U(1) \sim \mathbb{CP}^1$. The coset has the $U(1)$ symmetry: this exactly the $U(1)$ symmetry that emerged in our on-shell analysis.

At this point, the engaged reader may complain: but you said that the vanishing soft limits were associated with a shift symmetry $Z \to Z + c + \dots$\,!!!? This does not appear to be a symmetry of the Lagrangian \reef{CP1Lag}. However, a Lagrangian is not unique but can take a different form on field redefinitions; this cannot change the physical observables, the amplitudes. So the shift symmetry can be accompanied by a field redefinition and in fact the Lagrangian  \reef{CP1Lag} is invariant under infinitesimal shift symmetry
\be
  Z \to Z + c + \bar{c} \frac{a_1}{2 \Lambda^2} Z^2 \,,
  ~~~~~
  \bar{Z} \to \bar{Z} + \bar{c} + {c} \frac{a_1}{2 \Lambda^2} \bar{Z}^2 \,.
\ee
It is one of the appealing features of the on-shell amplitudes approach that it is independent of having to deal with redundancies such as those arising from field redefinitions (or gauge transformations). 

The very simple amplitude analysis shows that at the leading 2-derivative order, the model had to have $U(1)$ symmetry; in that sense it is {\em emergent}! Recognizing the leading-order model as  the $\mathbb{CP}^1 \sim SU(2)/U(1)$ NLSM, it is clear why the $U(1)$ had to be there. Next we discuss interactions beyond the leading order. 

\subsection{Beyond Leading Order}
\label{s:hdcorrections}
 In the $\mathbb{CP}^1$ NLSM, it is fairly easy to see that there is only one possible 2-derivative operator at 4-point: $Z \bar{Z} \partial_\mu Z \partial^\mu \bar{Z} = \tfrac{1}{4}(\partial_{\mu} Z^2) (\partial_{\mu} \bar{Z}^2)$. Any other way of arranging the derivatives on the four fields is equivalent to this using integration by parts and the leading order equations of motion (EOM) 
$\Box Z = \partial_\mu \partial^\mu Z = 0$ and $\Box \bar{Z} = 0$. But what about higher derivative corrections? There are many ways of sprinkling four derivatives on the four fields, but how many are actually independent under integration by parts and use of the EOM? And with $2k$-derivatives? This is very easy to answer using the on-shell amplitude methods as we now demonstrate. 

\vspace{2mm}
\noindent {\bf Higher Order Corrections.} 
A $2k$-derivative term generates contributions to the amplitudes at $O(p^{2k})$. So the question of the number of independent $2k$-derivative operators is simply rephrased as: how many Bose symmetric 2nd order polynomials in the Mandelstam variables are independent under the relations of momentum conservation (translates to integration by parts) and on-shellness (translates to EOM)? 

For the 4-point $U(1)$-conserving  $O(p^4)$ case we find two such independent polynomials,  
\be
  A_4^{O(p^4)}(Z \bar{Z} Z \bar{Z} ) =  \frac{a_1}{\Lambda^2} \, s_{13} 
  + a_2 \, s_{13}^2 + a_2' \, (s_{12}^2 +  s_{14}^2) + O(p^6)\,.
\ee
The two terms correspond to the two independent Lagrangian terms   $\partial_\mu Z \partial_\nu \bar{Z}  \partial^\mu  Z \partial^\nu   \bar{Z}$  
  and $Z \bar{Z} \partial_\mu \partial_\nu  Z \partial^\mu \partial^\nu   \bar{Z}$. 
  
For the cases of $2k$-derivative one similarly finds 
\be
  \begin{array}{r|cccccccccc}
    2k~&
       0 & 2 & 4& 6 & 8 & 10 & 12 \\
      \hline 
      \# ~\text{independent $\pa^{2k} Z^2 \bar{Z}^2$ operators } & 
       1& 1 & 2 & 2&  3 & 3 & 4
  \end{array}
\ee
As $k$ or $n$ grows, it gets harder to brute force determine the number of independent Mandelstam polynomials subject to the given constraints. However, there are powerful mathematical tools, such as the Gr\"obner basis, for solving such problems and they have indeed been applied for these purposes \cite{Beisert:2010jx}.

At 4-point, the matrix elements trivially satisfy the vanishing soft-limit condition, simply due to the special 3-particle kinematics of the resulting limit. So one has to analyze more carefully at 6-point level (and higher) whether cancellations can occur to ensure that the soft limit gives zero.

\vspace{3mm}
\noindent {\bf Lowest order $U(1)$-Violating Operators.}
At leading order, the model we consider has $U(1)$ symmetry, but at subleading orders, our formulation of the problem allows for $U(1)$-violating terms. We found in Section \ref{s:cp1bootnoU1} that at 4-point, the $U(1)$-violating amplitudes start at $O(p^4)$. The explicit matrix elements at this order are 
\be
\label{u1viol4pt}
  A_4(ZZZZ) = \frac{d_1}{\Lambda^4} \big(s_{12}^2+s_{13}^2+s_{23}^2\big)+ O(p^6)\,,
  ~~~~~
  A_4(ZZZ \bar{Z})  = \frac{d_2}{\Lambda^4} \big(s_{12}^2+s_{13}^2+s_{23}^2\big)+ O(p^6)\,,
\ee
and similarly for those with conjugate states. 

At 5-point, the all-$Z$ matrix element is non-vanishing at  $O(p^4)$, 
 \be
  \label{u1viol5pt}
  A_5(ZZZZZ) = \frac{d_3}{\Lambda^5}\sum_{i<j} s_{ij}^2 + O(p^6)\,,
\ee
however, this amplitude does not vanish in the soft limit. Likewise, $A_5(ZZZZ\bar{Z})$ and $A_5(ZZZ\bar{Z}\bar{Z})$ have 2 and 3 independent matrix elements, respectively, but no linear combination of them vanish in the soft limit. At $O(p^6)$, there are two independent terms in the 5-point amplitude with vanishing soft limit:
\be
   \begin{split}
   A_5(ZZZ\bar{Z}\bar{Z})
   = &\frac{e_1}{\Lambda^7} s_{45} \big[ (s_{14}+s_{15})^2+(s_{24}+s_{25})^2 + (s_{34}+s_{35})^2 - 2s_{45}^2\big] \\
   &+\frac{e_2}{\Lambda^7}   \big[  s_{145} s_{14} s_{15} + s_{245} s_{24} s_{25}
   +s_{345} s_{34} s_{35}\big]\,.
   \end{split}
\ee
Both  vanish trivially as $p_4$ or $p_5 \to 0$. When one of the $Z$ particles goes soft, say $p_1 \to 0$, the resulting 4-particle kinematics $p_2+p_3+p_4+p_5 = 0$ ensures that both  expressions in $[... ]$ vanish.  These $O(p^6)$ operators are subleading to the $O(p^4)$ $U(1)$-violating ones from \reef{u1viol4pt}. 
There are of course many other operators one can construct. These examples simply illustrate the amplitude-based method.

 \subsection{Postscript: Coset Story}
 \label{s:coset}
The context of the problem studied here is the low-energy physics of Goldstone bosons arising from spontaneous breaking of an internal symmetry group $G$ to a subgroup $H$. The number of Goldstone bosons is equal to the number of broken symmetry generators, i.e.~dim$(G/H)$=dim($G$)-dim($H$). The scalars `live' in the coset space $G/H$. When $G/H$ is a symmetric space and there are no cubic interactions,  it can be shown that the amplitudes of the Goldstone bosons vanish in the single scalar soft limit.

For specific symmetry breaking patterns $G \to H$, there are techniques for systematic construction of Lagrangians of the Goldstone modes \cite{Coleman:1969sm,Callan:1969sn,Volkov:1973vd}. But for more open-ended questions aimed at understanding the space of possible theories and any additional emergent symmetries they may have, the Lagrangian approach is limited. 

In our particular example with two real Goldstone modes, there are two obvious candidate theories: 1) $SU(2)$ broken to  $U(1)$, or 2) $U(1) \times U(1)$ completely broken. As we have seen, the vanishing soft limit criterion selects for the former. The example serves to illustrate how the bottom-up approach to construction of theories via amplitude can have symmetries emerge that were not part of the input assumptions. There are examples where the emergence of symmetries are perhaps more surprising. This is, for example, the case with supersymmetric extensions of the $\mathbb{CP}^1$ model. When constructed from the on-shell amplitudes approach, one finds \cite{Elvang:2018dco} that not only does the $\mathcal{N}=1$ supersymmetric $\mathbb{CP}^1$  model have the $U(1)$ symmetry under which the scalars $Z$ and $\bar{Z}$ are charged (and their fermions are uncharged), it also has a second global $U(1)$ symmetry under which the scalars and fermions in the same supermultiplet have the same charge.

\section{Technical: Conformal Bootstrap}
\label{s:confboot}
This section gives a more technical account of  the conformal bootstrap, offering some details that was left out in the introduction of the method given in Section \ref{s:introbootstrap}. We start with some CFT background, then discuss the bootstrap method. 

\subsection{CFT background}

As described in Section \ref{s:introbootstrap}, an operator $\mathcal{O}_\Delta(x)$ is characterized by its spin $s$ (how is transforms under the Lorentz group) and its scaling dimension $\Delta$ introduced via the homogenous scaling property \reef{defDelta}. Unitary enforces a lower bound on the possible value of $\Delta$ for given spin $s$.  For a scalar operator, ($s=0$) in $d$-dimensions, this bound is 
\be
  \label{Deltabound}
  \Delta \ge d/2 -1 \,.
\ee
 The bound is exactly saturated for a free scalar field, which has $\Delta =d/2 -1$.  
 
 The correlation functions $\< \mathcal{O}_1(x_1) \mathcal{O}_2(x_2)\ldots \>$, i.e.~the vacuum expectation values of strings of local operators at different space time locations $x_i$, have to respect Poincar\'e symmetry, in particular translation invariance means that they depend on the spacetime coordinates only via the differences $x_{ij}^\mu = (x_i -x_j)^\mu$. In particular, this means that a 1-point function $\< \mathcal{O}(x)\>$ must be a constant. By dimensionality, it therefore has to vanish in a CFT since a conformal theory has no dimensionful constants.  

One can use scale invariance to prove that a 2-point correlation function  
$\< \mathcal{O}_{1}(x_1) \mathcal{O}_{2}(x_2) \>$ 
 vanishes unless $\Delta_1 = \Delta_2$. Moreover, the form of the correlation function is completely fixed by symmetries and one can organize the operators such that the 2-point correlation functions of scalar operators take the form
\be
  \label{2ptcorr}
 \< \mathcal{O}_i(x_i) \mathcal{O}_{i}(x_j) \>  = \frac{\delta_{ij}~}{|x_{ij}|^{2\Delta}} \,,
\ee
where $|x_{ij}|^2 = x_{ij}^\mu x_{ij\mu}$. This expression has the correct scaling behavior under \reef{defDelta}.

Scale invariance fixes a 3-point correlation function up to a constant $\lambda_{ijk}$ as
\be
  \label{3ptcorr}
 \<\mathcal{O}_i(x_i) \mathcal{O}_j(x_j) \mathcal{O}_k(x_k) \>  
 = \frac{\lambda_{ijk}}{|x_{ij}|^{\Delta_i + \Delta_j - \Delta_k} |x_{ik}|^{\Delta_i + \Delta_k - \Delta_j} |x_{jk}|^{\Delta_j + \Delta_k - \Delta_i}} \,.
\ee
In a unitary theory, the $\lambda_{ijk}$'s are real. 

Now it feels like we are on a good roll with scale invariance determining almost everything for us, but the fun stops at 3-point. Or maybe we should say that the fun begins at 4-points? Starting with 4-point correlation functions, one can build conformal cross-ratios like 
\be
  \label{uv}
  u = \frac{|x_{12}|^2 |x_{34}|^2}{|x_{13}|^2 |x_{24}|^2}
  ~~~~~\text{and}~~~~
  v = \frac{ |x_{14}|^2 |x_{23}|^2}{|x_{13}|^2 |x_{24}|^2} \,,
\ee
which are scale-invariant. One can also check that they invariant under inversion, hence since conformal boosts are  inversion-translation-inversion, they are also conformally invariant. 
As far as scale invariance is concerned,  a 4-point correlation function can depend in arbitrary ways on $u$ and $v$; for example for four identical scalar operators with scaling dimension $\Delta$, the expression
\be
 \label{4ptA}
 \<\mathcal{O}(x_1) \mathcal{O}(x_2) \mathcal{O}(x_3)  \mathcal{O}(x_4) \>  
  = \frac{1}{|x_{12}|^{2\Delta}|x_{34}|^{2\Delta}} \, g(u,v) \,
\ee
has the correct scaling for any function $g$ of the conformal cross-ratios \reef{uv}.
We could also have exchanged $2 \lra 4$ and written 
\be
 \label{4ptB}
 \<\mathcal{O}(x_1) \mathcal{O}(x_2) \mathcal{O}(x_3)  \mathcal{O}(x_4) \>  
  = \frac{1}{|x_{14}|^{2\Delta}|x_{23}|^{2\Delta}} \, g(v,u) \,.
\ee
Note that $u \lra v$ under $2 \lra 4$, hence the exchange of the arguments in the function $g$. The two expressions \reef{4ptA} and \reef{4ptB}  must give rise to the same correlation function, so the function $g$ is in fact not completely arbitrary but must obey 
\be \label{gflip}
 g(u,v)  = \bigg(\frac{u}{v}\bigg)^{\Delta}\,g(v,u)  \,.
\ee
This constraint plays a role in the following. 

Let us now dive into a little more depth with the Operator Product Expansion (OPE) expansion than we did in Section \ref{s:introbootstrap}. 
When  $x^\mu$ is close to $y^\mu$, the product $\mathcal{O}_i(x) \mathcal{O}_j(y)$ of two local operators create a local fluctuation and as such it should be possible to describe it by some linear combination of local operators. In particular, in  the limit $x^\mu \to 0$, the OPE expansion is
\be
   \label{OPE}
   \mathcal{O}_i(x) \mathcal{O}_j(0) \sim \sum_{n} c_{ijn}(x) \, \mathcal{O}_n(0)\,.
\ee
 The OPE functions $c_{ijn}(x)$ depend on $x$ and can in general be expected to be divergent as $x\to 0$. Applied to the 3-point correlation function 
$\<\mathcal{O}_i(x_i) \mathcal{O}_j(0) \mathcal{O}_k(x_k) \>$ as  $x_i \to 0$, the OPE \reef{OPE} reduces it to a (sum of) simple 2-point correlators \reef{2ptcorr}. If we compare that result with the expression \reef{3ptcorr} with $x_j = 0$ and expand around $x_i = 0$, we find 
\be
   c_{ijk}(x) = \frac{\lambda_{ijk}}{|x|^{\Delta_i+\Delta_j-\Delta_k}} \Big (1 
   +  \alpha \, x^\mu \frac{\pa}{\pa x_k^\mu}
   +
   \ldots\Big)
\ee 
where the dots stand for terms with subleading powers in small $x$ with coefficients (like $\alpha$) that depend on the operator scaling dimensions $\Delta_i$,  $\Delta_j$, and  $\Delta_k$. This means that the OPE can  be written 
\be
   \label{OPE2}
   \mathcal{O}_i(x_i) \mathcal{O}_j(x_j) = \sum_{n} \lambda_{ijn} \,C_{ijn}(x_{ij},\tfrac{\pa}{\pa x_j}) \, \mathcal{O}_n(x_j)\,,
\ee
where the sum is over {\em primary operators} $\mathcal{O}_n$. One can think of primary operators as the operators that cannot be obtained as derivatives (descendants) of other operators.
Because of their appearance in \reef{OPE2}, the constants $\lambda_{ijk}$ in the 3-point correlation functions  \reef{3ptcorr} are called {OPE coefficients}.

One can apply the OPE to obtain expressions for higher-point correlation functions in terms of the so-called {\em CFT data}:
\begin{itemize}
\item A list $\{\Delta_i, R_i\}$ of all the operator scaling dimensions $\Delta_i$ and spin representations $R_i$ of the local primary operators of the theory.
\item A list of all OPE coefficients $\lambda_{ijk}$.
\end{itemize}
Not any list of $\{\Delta_i, R_i\}$ and OPE coefficients define a CFT; there are certain constraints that must be obeyed. At the core, the conformal bootstrap is about how to impose  such a constraint and the remarkable milage gained from it.

\subsection{Bootstrap}
Suppose we apply the OPE \reef{OPE2} twice in a 4-point correlation function to get
\be
  \label{4ptOPE12}
  \big\< 
  \mathcal{O}_1(x_1) \mathcal{O}_2(x_2) \mathcal{O}_3(x_3) \mathcal{O}_4(x_4) \big\> 
  = \!
  \sum_{ \mathcal{O}^a,  \mathcal{O}^b} 
  \lambda_{12a} 
  \lambda_{34b} \,
  C_{12a}(x_{12},\pa_2)
  C_{34b}(x_{34},\pa_4)
  \big\< \mathcal{O}^a(x_2) \mathcal{O}^b(x_4) \big\>.
\ee
The sum is over primary operators and $a, b$ are collective indices that include both operators and the index structure associated with their spin. 

Here we made a choice to pair 1 and 2 in the OPE and 3 with 4, but obviously there are 2 other possible choices. These three choices have to be equivalent since they are simply different representations of the same correlation function. The requirement of equivalence is very non-trivial and gives rise to the {\em crossing relations}, sometimes also called {\em OPE associativity}. Pictorially we can illustrate the crossing relations for the 12 pairing and the 14 pairing as 
\be
   \sum_\mathcal{O} \lambda_{12\mathcal{O}} \lambda_{34\mathcal{O}}
   \raisebox{-0.6cm}{
    \begin{tikzpicture}[scale=0.4, line width=1 pt]
    	\draw (-1,1)--(0,0);
	\draw (-1,-1)--(0,0);
	\draw (0,0)--(2,0);
	\draw (2,0)--(3,1);
	\draw (2,0)--(3,-1);
	\node at (-1.4,1) {$1$};
	\node at (-1.4,-1) {$2$};
	\node at (3.4,-1) {$3$};
	\node at (3.4,1) {$4$};
	\node at (1.2,0.55) {$\mathcal{O}$};
    \end{tikzpicture}
    }
   ~~=~~
   \sum_{\mathcal{O}'} \lambda_{14\mathcal{O}'} \lambda_{23\mathcal{O}'}
   \hspace{-0.3cm}
      \raisebox{-1.cm}{
    \begin{tikzpicture}[scale=0.4, line width=1 pt]
    	\draw (-1,2)--(0,1);
	\draw (1,2)--(0,1);
	\draw (0,-1)--(0,1);
	\draw (-1,-2)--(0,-1);
	\draw (1,-2)--(0,-1);
	\node at (-1.4,2) {$1$};
	\node at (-1.4,-2) {$2$};
	\node at (1.45,-2) {$3$};
	\node at (1.45,2) {$4$};
	\node at (0.8,0.1) {$\mathcal{O}'$};
    \end{tikzpicture}
    }
  \,.      
\ee

It is not hard to see that if the 4-point correlation functions satisfy crossing relations, then it is also true of the $n$-point functions. For the purpose of this presentation, we consider a CFT to be defined as a set of CFT data ($\{\Delta_i, R_i\}$ and $\lambda_{ijk}$'s) that satisfy the crossing relations for 4-point correlators.\footnote{~Other constraints may also be useful or needed, such as modular invariance in 2d or constraints arising from supersymmetry or the presence of boundaries, interfaces, and line operators \cite{Simmons-Duffin:2016gjk}.} 

As described in Section \ref{s:introbootstrap}, the idea of the {\em conformal bootstrap} is to take a set of CFT data of a putative CFT with specified symmetries and apply the constraints of the crossing relations to find out if such a CFT may exist.

Take all four operators in the 4-point correlator to be identical scalar operators $\mathcal{O}$ with scaling dimension $\Delta$. The expression \reef{4ptOPE12} can then be written 
\be
    \big\< 
  \mathcal{O}(x_1) \mathcal{O}(x_2) \mathcal{O}(x_3) \mathcal{O}(x_4) \big\> 
  = 
  \frac{1}{|x_{12}|^{2\Delta}|x_{34}|^{2\Delta}} 
  \sum_{ \mathcal{O}^a} 
  \lambda_{\mathcal{O}\mathcal{O}a}^2
  \,
  g_{\Delta_a,\ell_a}(u,v)\,,
\ee
where we have introduced the {\em conformal blocks}
\be
  g_{\Delta_a,\ell_a}(u,v)
  =
  |x_{12}|^{2\Delta}|x_{34}|^{2\Delta}
  \,C_{12a}(x_{12},\pa_2)
  \,C_{34b}(x_{34},\pa_4)
  \,
  \frac{Y_{ab}}{|x_{24}|^{2\Delta}}
\ee
using
\be
  \big\< \mathcal{O}_a(x_2) \mathcal{O}_b(x_4) \big\>
  =
  \frac{Y_{ab}(x_{24})}{|x_{24}|^{2\Delta}}
\ee
in which  $Y^{ab}$ captures the appropriate index structure for operators with non-vanishing spin and is simply $\delta^{ab}$ for scalar operators. (For the case here, only operators with even spin appear in the OPE.)  Comparing with \reef{4ptA}, we see that the OPE has decomposed the function $g(u,v)$ as 
\be
  g(u,v)
  = 
  \sum_{ \mathcal{O}^a} 
  \lambda_{\mathcal{O}\mathcal{O}a}^2~
  g_{\Delta_a,\ell_a}(u,v) \,.
\ee
This sum over primary operators is called the conformal block decomposition. 

Now we saw already in the previous subsection that the different ways of decomposing the 4-point correlation function required the function $g$ to satisfy \reef{gflip}, i.e.~$v^{\Delta}  g(u,v) - u^{\Delta}  g(v,u)  = 0$. This means that the crossing relation becomes the mathematical requirement 
\be
  \label{crossing1}
  \sum_{ \mathcal{O}^a} 
  \lambda_{\mathcal{O}\mathcal{O}a}^2~
  \Big(
  \underbrace{v^{\Delta} g_{\Delta_a,\ell_a}(u,v)
  -
  u^{\Delta} g_{\Delta_a,\ell_a}(v,u)}_{F^{\Delta}_{\Delta_a,\ell_a}(u,v)}
  \Big) 
  ~=~0
  \,.
\ee

The conformal blocks $g_{\Delta_a,\ell_a}(u,v)$ are fixed by conformal symmetry, for example via a differential equation derived from the quadratic Casimir (or by series expansion or by recursion relations). In 2d and 4d, they can be expressed in terms of hypergeometric functions ${}_2F_1$. Hence, the functional form of $F^{\Delta}_{\Delta_a,\ell_a}(u,v)$ is known, and the only unknown ingredients in \reef{crossing1} are the scaling dimensions and the OPE coefficients. Thus for given input CFT data, the crossing relations \reef{crossing1} is a consistency condition.  It is the central work-horse of the conformal bootstrap. 

We mentioned in \reef{schm} that the bootstrap equation \reef{crossing1} has a geometric interpretation: for a given list of vectors $\vec{v}_{\sigma\sigma\mathcal{O}}$, which we now identify as $F^{\Delta}_{\Delta_a,\ell_a}(u,v)$, does there exist real non-negative coefficients\footnote{ The OPE coefficients were called $c_{ijk}$ in Section \ref{s:introbootstrap} because we glossed over the dependence on the spacetime coordinates.} $\lambda_{\mathcal{O}\mathcal{O}a}^2$ such that the linear combination \reef{crossing1} vanishes? If one can show that for given assumptions on the conformal dimensions, all $F^{\Delta}_{\Delta_a,\ell_a}(u,v)$'s lie on one side of some plane then no CFT can exist with that data because it could never satisfy the crossing relations.

The basic approach in the (numerical) bootstrap can be described algorithmically:
\begin{itemize}
\item Make assumptions about the spectrum of the lowest dimension operators of a putative CFT: their scaling dimensions and spin. 
\item Test the crossing relation. Numerically, can be done by searching for a functional $\alpha$ such that $\alpha (F^{\Delta}_{\Delta_a,\ell_a}(u,v)) \ge 0$.
\end{itemize}
 If such an $\alpha$ exists, then the crossing relations can never be satisfied and no such CFT can exist. This means that the bootstrap algorithm is especially powerful for ruling theories out. If no $\alpha$ is found, it is not prove that a theory exists, it is at best a ``maybe".  
 Scanning through the space of possible scaling dimensions of the lowest dimension operators numerically using semidefinite programming techniques has resulted in powerful bounds on existence of CFTs in diverse dimensions, as we described for the 3d Ising model and other applications in Section \ref{s:introbootstrap}. We finish with one other example.

\subsection{Example: Infinite Number of Primary Operators}

As an example (borrowed from \cite{Rattazzi:2008pe,Rychkov:2009ij}, see also the review \cite{Simmons-Duffin:2016gjk}) of analytic bootstrap, we ask if there exist any 4d CFTs with a finite number of primary operators?  To answer this question, introduce complex variables $z$ and $\bar{z}$ such that the conformal cross-ratios \reef{uv} are 
\be
  u = z\bar{z} ~~~\text{and}~~~ v = (1-z)(1-\bar{z})\,.
\ee
Using the representation of the conformal blocks in terms of hypergeometric functions one can show that in the limit $z \to 0$, taken along the line $\bar{z} = z$, one has
\be
  g_{\Delta_a,\ell_a}(u,v) \to z^\Delta + \ldots
  ~~~~\text{and}~~~~
g_{\Delta_a,\ell_a}(v,u) \to \log(z) + \ldots \,.
\ee
This means that $g(u,v)$ is dominated by the operator with the smallest scaling dimension, namely the identity operator which has $\Delta=0$, so that
\be
  g(u,v) =\lambda_{\mathcal{O}\mathcal{O}1}^2 \cdot 1 + \ldots\,.
\ee
On the other hand, as $z \to 0$ along the line $\bar{z} = z$, the other side of the crossing relation \reef{gflip} behaves as
\be
   \bigg(\frac{u}{v}\bigg)^{\Delta}  g(v,u)  
  \to z^\Delta \sum_{\mathcal{O}^a} \lambda_{\mathcal{O}\mathcal{O}a}^2 \log(z) + \ldots
\ee
Clearly this vanishes unless there are infinitely many terms in the sum. So in order for crossing to hold in the form \reef{gflip},  {\em there must be an infinite number of primary operators in any 4d CFT.}

\section{Concluding Remarks}
\label{s:conclude}

The amplitudes program and the conformal bootstrap share a common `philosophy': at the center of the explorations are the physical observables, the on-shell amplitudes and the correlation functions, respectively. The overlap goes beyond that, for example in the increasing use of common tools. Joint workshops, like the two summer programs on the Conformal Bootstrap and Amplitudes in 2015 and 2019 at the Aspen Center for Physics contribute to the increased communication between the communities of researchers. And that will hopefully continue in the future. 

\section*{Acknowledgements}

I would like to thank Gordon Baym for the invitation to write up my Aspen Colloquium. I am grateful to Huan-Hang Chi, Aidan Herderschee, Callum Jones, Matt Mitchell, and Shruti Paranjape 
for discussions and detailed feedback on the manuscript.  I would like to thank  Yu-tin Huang, David Poland, Rafael Porto, Silviu Pufu, and Slava Rychkov for helpful comments and corrections. 
This work was initiated at the Aspen Center for Physics which is supported by National Science Foundation grant PHY-1607611. The author is supported in part by the US Department of Energy under Grant No.~DE-SC0007859.

\section*{References}
\bibliographystyle{JHEP}
\bibliography{amsbootBIB.bib}


\end{document}